\documentclass[aps,prb,longbilbiography,twocolumn]{revtex4-1}
\usepackage{amsfonts}
\usepackage{amsmath,latexsym}
\usepackage{psfrag}
\usepackage{amssymb,amsmath,amsthm}
\usepackage[dvips]{graphicx}
\usepackage{verbatim} %
\usepackage{color}
\usepackage{braket}
\newcommand{\beq}{\begin{equation}}
\newcommand{\eeq}{\end{equation}}
\newcommand{\bea}{\begin{eqnarray}}
\newcommand{\eea}{\end{eqnarray}}
\usepackage{bm}
\usepackage{color}

\newcommand{\bma}{\left(\begin{matrix}}
\newcommand{\ema}{\end{matrix}\right)}
\def\build#1_#2{\mathrel{\mathop{#1}\limits_{#2}}}
\definecolor{pink}{rgb}{1,0.5,0.5}
\definecolor{violet}{rgb}{1,0,1} 
\definecolor{red}{rgb}{1,0,0}
\definecolor{yellow}{rgb}{0.7,1,0}
\definecolor{orange}{rgb}{1,0.5,0}
\definecolor{white}{rgb}{1,1,1}
\definecolor{blue}{rgb}{0,0,1}
\definecolor{cyan}{rgb}{0,1,1}

\def\S{{\mathcal S}}

\def\T{{\mathcal T}}

\def\b{{\bf b}}

\def\k{{\bf k}}

\def\x{{\bf x}}
\def\y{{\bf y}}
\def\z{{\bf z}}
\def\u{{\bf u}}
\def\v{{\bf v}}
\def\e{\bm{ \varepsilon}}

\begin{document}

\begin{abstract}
A canonical quantization procedure is applied to elastic waves interacting with pinned dislocation segments (``strings'') {of length $L$} via the Peach-Koehler force. {The interaction Hamiltonian, derived from an action principle that classically generates the Peach-Koehler force, is a power series of creation and annihilation operators. The leading term is quadratic, and keeping only this term the observable quantities of scattering processes are computed to all orders in perturbation theory. The resulting theory is characterized by the magnitude of $kL$, with $k$ the wavenumber of an incident phonon. The theory is solved for arbitrary $kL$, and different limits are explored. A significant result at this level is the scattering cross section for phonons by dislocation segments. As a function of frequency, this cross section has a  much richer structure than the linear-in-frequency behavior that is inferred from scattering by an infinite, static, dislocation. The rate of spontaneous phonon emission by an excited dislocation is computed as well.  When many dislocations are present, an effective mass operator is computed in the Weak and Independent Scattering Approximation. The contribution of the cubic terms is computed to leading order in perturbation theory. They allow for a comparison of the scattering of a phonon by a string and the three-phonon scattering, as well as studying the dependence of scattering amplitudes on the temperature of the solid. It is concluded that the effect of dislocations will dominate for relatively modest dislocation densities. Finally, the full power series of the interaction Hamiltonian is considered. The effects of quantum corrections, i.e., contributions proportional to Planck's constant, are estimated, and found to be controlled by another wavenumber-dependent parameter $k d_q$, where $d_q$ is a length proportional to $\sqrt{\hbar}$. The possibility of using the results of this paper in the study of the phononic thermal properties of two- and three-dimensional materials is noted and discussed.}

\end{abstract}

\title{Scattering of phonons by quantum dislocations segments in an elastic continuum}

\author{Fernando Lund$^{1,2}$ and Bruno Scheihing H.$^1$}

\affiliation{\mbox{$^1$Departamento de F\'\i sica and $^2$CIMAT, Facultad de Ciencias
F\'\i sicas y Matem\'aticas, Universidad de Chile, Santiago, Chile} }

\date{ \today}

\maketitle

\section{Introduction} 
The interaction of phonons with dislocations has long been a subject of concern. Thermal transport, particularly in insulators, and thermoelectric materials are two overlapping perspectives  where this concern is of central importance. In a classic paper, Klemens \cite{Klemens1955} computed the scattering of low frequency (so that dispersion effects could be neglected) phonons by a number of static imperfections, including infinitely long, straight dislocations. {This configuration was also considered by Carruthers \cite{Carruthers1959}, who considered a modified version of the potential energy.} One important insight that came out of this very simplified approach is that dislocations should be far more efficient scattering phonons that travel perpendicular, rather than parallel, to the dislocation line. To this day, these works appear to be the main source of information to analyze the role of dislocations in thermal transport experimental data \cite{Phillpot2013,Cahill2014,Feng2014}.

Quite recently, it has become possible to isolate the contribution of dislocations to thermal transport: Kim et al. \cite{Kim2015,Kim2017} obtained a significant lowering of thermal conductivity in Bi$_{0.5}$Sb$_{1.5}$Te$_3$ through the introduction of dense dislocation arrays within grain boundaries. Sun et al. \cite{Sun2018} have measured a significant thermal transport anisotropy in micron thick, single crystal InN films with  highly oriented dislocation arrays. Xie et al. \cite{Xie2018} report a 50\% reduction in lattice thermal conductivity due to dislocations in Si-based thermoelectric composites. Chen et al. \cite{Chen2017} have obtained very low values of lattice thermal conductivity in Pb$_{1-x}$Sb$_{2x/3}$Se solid solutions, by way of generating dense arrays of uniformly distributed dislocations in the grains. { Zhou et al. \cite{Zhou2018} have rationally included point defects, vacancy driven dense dislocations and nanoprecipitates in a n-type PbSe thermoelectric material to obtain a lattice thermal conductivity approaching that of the amorphous limit.} There clearly is a very real possibility of engineering defects, particularly dislocations, to obtain desired thermal performances, and the theoretical framework currently available is not enough to rationalize these measurements.

A dislocation that is hit by a phonon will typically respond by bowing out, oscillating, and re-radiating secondary phonons. This process can lead to resonances which may affect the scattering cross section, and the resulting relaxation time, considerably. At first sight, one could think that resonance would be effective only for phonons whose wavelength is comparable to the distance between pinning points for a pinned dislocation, say tens of nanometers, and hence not relevant at most working temperatures when the dominant phonon wavelength is much shorter. A closer look, however, reveals that, thinking of the dislocation as an elastic string, there will be resonant scattering not only when the phonon frequency coincides with the fundamental mode of vibration of the string, but also when it coincides with any of the higher order modes of vibration. This dynamical mechanism was tackled by Ninomiya \cite{Ninomiya1968,Ninomiya1969} for an infinitely long dislocation line that can oscillate around a straight equilibrium position, coupled to phonons through the kinetic energy terms but not through the potential energy terms. Building on the work of Ninomiya \cite{Ninomiya1968,Ninomiya1969}, Li et al. \cite{Li2017a,Li2017b,Li2018}  have recently developed a theory of phonons, and electrons, in interaction with the quantum normal modes (termed ``dislons'') of an infinitely long straight dislocation. The question naturally arises: What is is the quantum description of dislocation segments of finite length interacting with phonons? In this paper we address this issue.

The interaction of classical elastic waves with dislocations segments of finite length has been a topic of active research for decades \cite{Granato1956a,Granato1956b}. A comparatively recent result is the computation of the scattering cross section for classical elastic waves by pinned dislocation segments \cite{Maurel2005a}, as well as by prismatic dislocation loops \cite{Natalia2009}. Use of the formalism developed therein, together with multiple scattering theory, has enabled the computation of an effective, complex, index of refraction for the propagation of coherent elastic waves in a medium filled with randomly distributed and oriented, pinned dislocation segments \cite{Maurel2005b,Churochkin2016}. In turn, this has led to novel characterization tools to describe the plasticity of metals and alloys \cite{Mujica2012,Salinas2017,Espinoza2018}. 

The quantum theory of elastic waves in interaction with dislocations has been far less studied. In addition to the {analytical} work of Ninomiya \cite{Ninomiya1968,Ninomiya1969} and Li et al. \cite{Li2017a,Li2017b,Li2018} already mentioned,  Wang et al. \cite{Wang2017} have performed a {numerical} ab-initio calculation of phonon scattering by a dipole of edge dislocations in silicon. They infer, from their calculation, that high but realistic dislocation concentrations can significantly influence thermal conductivity at room temperature and above. Also, their results are qualitatively at variance with the classic ones \cite{Klemens1955,Carruthers1959}

This paper is organized as follows: Section \ref{Sec:classical_action} has a brief review of the classical theory that we shall quantize. Section \ref{canonical_quant} describes the canonical quantization of the free theory. {The interaction term involves polynomial interactions between phonons and quantum strings. Section \ref{sec:quad} solves the quantum theory retaining only the quadratic interactions but to all orders in perturbation theory, which turns out to be equivalent to have a solution for arbitrary phonon wavelengths: short, comparable, and long compared to string length $L$. Sections \ref{lowest_o} and \ref{all_o} provide the scattering cross section for phonons by dislocations, while Section \ref{mass_op} gives the mass operator when many dislocations are present in the weak and independent scattering approximation. Section \ref{sec:cubic} computes the contribution of cubic terms in the interaction, to leading order in perturbation theory. This allows for a comparison of phonon-dislocation scattering with three-phonon processes, including the effect of temperature. Finally, Section \ref{sec:Quant} solves the theory for the full polynomial interaction, to leading order in perturbation theory in which a characteristic length proportional to the square root of Planck's constant emerges. Section \ref{discussion} has concluding remarks. A number of technical computations are provided in four Appendices.}

\section{Classical action}
\label{Sec:classical_action}
In this work we work out the quantum theory of oscillating dislocation segments, of length $L$, in interaction with elastic waves in three dimensions.

\begin{figure}[h]
\centering
\includegraphics[width=0.3\textwidth]{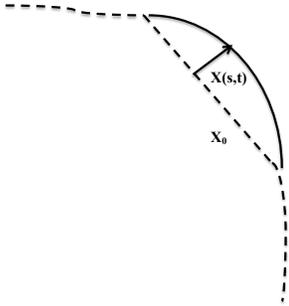}
\caption{The dislocation line has a static part ${\mathbf X}_0$, a portion of which is a straight line segment. Oscillations about this straight line segment are described by ${\mathbf X} (s,t)$, where $s$ is a position parameter and $t$ is time.}
\label{Fig1}
\end{figure}

{We consider a homogeneous, isotropic, elastic, continuum solid of density $\rho$ and elastic constants $c_{pqmr}, (p,q = 1,2,3)$ within which there is a string-like dislocation line. The variables describing the solid are the displacements  ${\mathbf u (\mathbf x,t)}$, at time $t$, of a point whose equilibrium position is ${\mathbf x}$.  In addition, there is a string described by a vector ${\mathbf X} (s,t)$, where $0<s<L$ is a position parameter along the string whose ends are fixed. The fact that the string is a dislocation is implemented through the fact that the displacements ${\mathbf u (\mathbf x,t)}$ are multivalued functions: They have a discontinuity equal to the Burgers vector $\b$ when crossing a surface, part of whose boundary includes the string with fixed ends. In addition to this geometrical fact, the coupling between elastic displacements and elastic string is given by standard conservation of energy and momentum arguments. When dislocation velocities are small compared to the speed of sound, an assumption we shall make throughout this work, this leads to the well-known Peach-Koehler force \cite{PeachKoehler1950}.  In the time-dependent case, and for string velocities small compared to the speed of sound, the dynamics is described by the following classical action \cite{Lund1988}:}

\beq
S= S_{\rm ph} + S_{\rm string} + S_{\rm int} + S_0
\label{actionFL}
\eeq
where 
\bea
S_{\rm ph} &=& \frac{1}{2} \int \! dt \! \int \! d^3 x  \! \left( \rho \dot{\mathbf u}^2 - c_{pqmr} \frac{\partial u_m}{\partial x_q} \frac{\partial u_p}{\partial x_r} \right) + \cdots \\
S_{\rm string} &=& \frac{1}{2} \int \! dt \! \int_0^L \!\!\! ds  \! \left( m \dot{\mathbf X}^2 - \Gamma {\mathbf X}'^2 \right) + \cdots \\
S_{\rm int} &=& - b_i \int \! dt \! \int_{\delta \mathcal{S}} dS^j \sigma_{ij}.
\eea
where the ellipses ``$\cdots $'' refer to higher order terms in the phonon or string actions, and $S_0$ involves the interaction of the elastic displacements with a static dislocation, part of whose boundary is the straight line ${\mathbf X}_0$ (Figure \ref{Fig1}). This is the interaction that was considered by Klemens \cite{Klemens1955} when ${\mathbf X}_0$ is an infinite line and will not be discussed further in this paper.

Let us start by describing the phonon action $S_{\rm ph}$: Clearly, $S_{\rm ph}$ describes elastic waves in an elastic continuum, wherein the quadratic terms will lead to free phonons and the higher order terms will lead to phonon-phonon interactions. {In the isotropic case} $c_{pqmr} = \lambda \delta_{pq} \delta_{mr} + \mu (\delta_{pm} \delta_{qr} + \delta_{pr} \delta_{mq} )$ {where $\lambda$ and $\mu$ are the Lam\'e constants.}

Classically, and in the absence of the interaction term, the free phonon theory has simple solutions in terms of plane waves for the elastic displacement (phonons) and normal modes for the pinned string. In the case of phonons, they are characterized by two different modes of propagation: transversal waves of speed $c_T = \sqrt{\mu/\rho}$ with two allowed polarizations, and longitudinal waves of speed $c_L = \sqrt{(\lambda+2\mu)/\rho}$ with one polarization. For future usage, we define $\gamma \equiv c_L/c_T > 1$.

On the other hand, we have a string described by a vector ${\mathbf X} (s,t)$, where $0<s<L$ is a position parameter along the string. We consider small deviations from a straight equilibrium position ${\mathbf X}_0$, the ends of which are pinned to their positions. $S_{\rm string}$ describes oscillations (normal modes) of an elastic string of finite length with fixed ends; higher order terms describe anharmonic effects on these oscillations, which we will not address in this work. The parameters $m$ and $\Gamma$ characterize the dislocation segment. {In this work we shall consider segments of edge dislocations only, in which case} they may be written in terms of the Burgers vector $\b$ as \cite{Lund1988}
\beq
m = \frac{\rho b^2}{4\pi} (1 + \gamma^{-4}) \ln (\delta/\delta_0)
\eeq
where $\delta$, $\delta_0$ are long- and short-distance cutoff lengths, and
\beq
\Gamma = \frac{\mu b^2}{2\pi} (1 - \gamma^{-2}) \ln (\delta/\delta_0).
\eeq
$m$ defines a mass per unit of length and $\Gamma$ the line tension. 

In the same manner as classically the phonons have plane waves solutions, this term of the action has oscillatory solutions that may be expanded in Fourier series ${\rm Re}\left\{ \sum_n a_n e^{- i\omega_n t} \sin(n \pi s/L) \right\}$, where $\omega_n = \frac{n\pi}{L}\sqrt{\frac{\Gamma}{m}}$ is the frequency of each normal mode. We take the string to have one degree of freedom (i.e. one direction orthogonal to its equilibrium position over which to oscillate) defined by the direction of the Burgers vector $\b$, the glide plane. For most of our results, the generalization to more directions of oscillation is straightforward.

Finally, $S_{\rm int}$ describes the interaction between these two sectors. It is given by the well known Peach Koehler force: $b_i$ is the $i$-th component of the Burgers vector, $\sigma_{ij}$ is the elastic stress tensor, {evaluated at the current position of the dislocation line,} and the surface $\delta \S$ describes the region bounded by the string and its equilibrium position. {As mentioned in the Introduction,} the classical scattering of phonons by dislocations using this interaction term has already been dealt with \cite{Maurel2005b,Churochkin2016}.

Having this realized, we now turn to quantize the theory by introducing canonical commutation relations.

\section{Canonical quantization of the free fields}
\label{canonical_quant}

We commence this section by implementing the correspondence of the Poisson brackets to commutators $\{ \cdot , \cdot \} \to -\frac{i}{\hbar} [ \cdot , \cdot]$. According to standard practice \cite{Srednicki2007}, the mode coefficients of the classical solutions are promoted to creation and annihilation operators, in terms of which we may write the displacement field as
\bea
\u (\x,t) &=& \sqrt{\frac{\hbar}{\rho}} \! \int \! \frac{d^3k}{(2\pi)^3} \!  \sum_{\iota \in \{{\rm pol.}\}}  \left[ \frac{\e_{\iota}^*(\k) a_{\iota}(\k) e^{ i\k \cdot \x -i\omega_{\iota}\! (\k)t}}{\sqrt{2 \omega_{\iota}\! (\k) }} \right. \nonumber \\  
&& \hspace{5em} \left.
 + \frac{\e_{\iota}(\k) a_{\iota}^{\dagger}(\k) e^{ - i\k \cdot \x +i\omega_{\iota}\! (\k)t}}{\sqrt{2 \omega_{\iota}\! (\k) }}   \right]
 \label{qdispl}
\eea
and the string displacement as
\beq
X(s,t) = \sqrt{\frac{\hbar}{mL}} \sum_{n=1}^{n_D} \left( \frac{a_{n} e^{-i\omega_n t}}{\sqrt{\omega_n}} +  \frac{a_{n}^{\dagger} e^{i\omega_n t}}{\sqrt{\omega_n}}   \right) \sin \! \left(\frac{n\pi s}{L} \right),
\eeq
where $n_D$ is a cutoff for the string wavenumber, as an oscillating string with the action $S_{\rm string}$ can only represent the physical setting down to a given length where the idealization breaks down. When comparing with experiments, if possible, this number should be left as a free parameter to be fixed by the actual results. In condensed matter a natural cut-off is provided by the Debye frequency.

Now we proceed to impose canonical commutation relations:
\bea
[u_i(\x,t), \rho {\dot u}_j(\y,t)] &=& i \hbar \delta_{ij} \delta^{(3)}(\x - \y), \\   
{[}X(s,t), m {\dot X}(s',t){]}  &=& i \hbar \delta(s-s'), \\
{[}u_i(\x,t), u_j(\y,t){]} = 0, & & {[}X(s,t), X(s',t){]} = 0,
\eea
which fully define the quantum theory in the non-interacting case. These relations in turn require
\beq
[a_{\iota}(\k), a_{\iota'}(\k')] = [a_{\iota}^{\dagger}(\k), a_{\iota'}^{\dagger}(\k')] = 0,
\eeq
\beq
[a_{\iota}(\k), a_{\iota'}^{\dagger}(\k)] = (2\pi)^3 \delta^{(3)}(\k - \k') \delta_{\iota \iota'}, 
\eeq
\beq
[a_{n}, a_{m}] = [a_{n}^{\dagger}, a_{m}^{\dagger}] = 0 \, , \,\,\,\, [a_{n},  a_{m}^{\dagger}] = \delta_{nm}.
\eeq
In the preceding expressions, $\iota$ is an index that runs over the possible polarizations for the phonons: two transversal modes and a longitudinal mode. The corresponding frequencies satisfy $\omega_{\iota}(\k) = c_\iota k,$ where ${\iota} = L, T$ for the phonons (denoting longitudinal and transverse propagation respectively), and $\omega_n = \frac{n\pi}{L}\sqrt{\frac{\Gamma}{m}}$ for the string. Finally, $\e_\iota(\k)$ represents the polarization vector associated to each mode of propagation. These vectors satisfy
\bea
\e_{\iota}(\k) \cdot \e_{\iota'}^*(\k) &=& \delta_{\iota \iota'}, \\ 
\sum_{\iota = T_1, T_2} \varepsilon^*_{\iota}(\k)_i  \varepsilon_{\iota}(\k)_j &=& \delta_{ij} - \frac{k_i k_j}{k^2}, \\
\varepsilon^*_{L}(\k)_i  \varepsilon_{L}(\k)_j &=&  \frac{k_i k_j}{k^2},
\eea
where $T_1,T_2$ denote the two transversal polarizations and $L$ the longitudinal polarization. The Hamiltonian of the free theory is
\beq
H = H_{\rm ph} + H_{\rm string}
\eeq
with phonon and string terms given, respectively, by
\bea
 H_{\rm ph} & = & \int \! \frac{d^3k}{(2\pi)^3} \!  \sum_{\iota \in \{{\rm pol.}\}} \hbar \omega_{\iota}(\k) a_{\iota}^{\dagger}(\k) a_{\iota}(\k)\\
 H_{\rm string} & = &   \sum_{n=1}^{n_D} \hbar \omega_n  a_{n}^{\dagger} a_{n}
 \eea

In characterizing the free theory, a fundamental object is the two-point function, or more commonly known as the \textit{propagator}. Let $\text{T}$ be the time-ordering symbol, instructing operators evaluated at a later time to be placed at the left, and let $\ket{0}$ be the vacuum state of the quantum mechanical system, with no excitations of the elastic displacements nor the string. For the dislocation, it reads
\begin{equation}
\begin{split}
\Delta(s,s',t-t') \equiv & \braket{0 | \text{T} X(s,t) X(s',t') | 0} \\ = & \frac{\hbar}{mL} \sum_{n=1}^{n_D} \frac{e^{-i \omega_n |t-t'| }}{\omega_n} \sin \! \left(\frac{n\pi s}{L} \right) \sin \! \left(\frac{n\pi s'}{L} \right).
\end{split}
\end{equation}
Even though we can do the same for the elastic displacement field, it turns out to be more useful for subsequent computations to write down the propagator for its spatial derivative:
\begin{equation}
\begin{split}
\Delta_{iji'j'} & (\x - \x', t- t') \\ \equiv & \braket{0 | \text{T} \frac{\partial u_i}{\partial x_j}(\x,t) \frac{\partial u_{i'}}{\partial x_{j'}}(\x',t') | 0} \\ = & \frac{\hbar}{\rho}  \int \frac{d^3k}{(2\pi)^3} k_j k_{j'} \left[ \left(\delta_{ii'} - \frac{k_i k_{i'}}{k^2} \right)\frac{e^{-i \omega_T(\k) |t - t'|} }{2 \omega_T(\k)} \right. \\
 & \,\,\,\,\,\,\,\,\,\,\,\,\,\,\,\,\,\,\,\,\,\,\,\,\,\,\,\,\,\,\,\,\,\,\,\,\,\,\,\,\,\,\, \left. + \frac{k_i k_{i'}}{k^2} \frac{e^{-i \omega_L(\k) |t - t'|} }{2 \omega_L(\k)} \right] e^{i \k \cdot (\x - \x')}.
\end{split}
\end{equation}

The reason behind writing down time-ordered quantities is that when we compute scattering amplitudes in the interacting theory, the $S$-matrix will be given by \cite{Srednicki2007}
\beq
\braket{\Psi_{\rm out} | \text{T} \exp \left[ - \frac{i}{\hbar} \int_{-\infty}^{\infty} H_I(t) dt \right] \! |\Psi_{\rm in} }
\eeq
where $H_I$ is the quantum mechanical interaction picture Hamiltonian operator, and thus the time-ordered quantities will have a central role.

In the sections that follow, we will take the string to have its equilibrium position along the $\hat{z} = \hat{e}_3$ axis, and the burgers vector to be written as $\b = b \hat{e}_1 = b \hat{x}$.

\section{The quadratic interactions: a single string} \label{sec:quad}

The lowest order interaction, which classically means to consider small strains, and small string excursions, away from the equilibrium position, is given by
\beq
S_{\rm int} = -\mu b \int dt   \int_0^L ds \, {\bf M}_{kl} \frac{\partial u_k}{\partial x_l}(\x_0+(0,0,s),t) X(s,t) 
\eeq
which is quadratic in the fluctuations. With our choice of coordinates, ${\bf M}_{kl}= (\hat{e}_1)_k (\hat{e}_2)_l +  (\hat{e}_2)_k (\hat{e}_1)_l $.  This interaction will give rise to the scattering of phonons by the strings, which is described by
\beq \label{ph-ph-a}
\langle f|i\rangle = \braket{0| a_{\iota}(\k) \text{T} \exp \left[ - \frac{i}{\hbar} \int_{-\infty}^{\infty} H_I(t) dt \right] a_{\iota'}^{\dagger}(\k')  |0},
\eeq
where
\beq \label{quad-hamiltonian}
H_I(t) = \mu b   \int_0^L ds \, {\bf M}_{kl} \frac{\partial u^{I}_k}{\partial x_l}(\x_0+(0,0,s),t) X^{I}(s,t).
\eeq
For completeness, we also write down the Hamiltonian in the Heisenberg picture (where it is naturally constant) in terms of creation and annihilation operators:
\beq
\begin{split}
H_{\rm int} =& \, \frac{\hbar}{\sqrt{ m L}} \int \frac{d^3k}{(2\pi)^3} \sum_{n=1}^{n_D} \sum_{\iota \in \{ {\rm pol.} \} } \left( \frac{a_n + a_n^{\dagger}}{\sqrt{\omega_n}} \right) \\
& \times \left(\frac{i n \pi E(\k;\iota) a_\iota(\k)}{L} \frac{e^{iL(\k \cdot \hat{e}_3) - i n \pi} - 1}{(\k \cdot \hat{e}_3)^2 - (n\pi/L)^2} + {\rm h. c.} \right)
\end{split}
\eeq
where h.c. stands for hermitian conjugate. Furthermore, we have defined $\omega \equiv \omega_\iota(\k) = \omega_{\iota'}(\k')$,
\beq
E(\k;\iota) \equiv \frac{\mu b}{\sqrt{2\rho \omega_{\iota}(\k) }} k_l {\bf M}_{kl} \varepsilon_\iota(\k)_k   e^{i \x_0 \cdot \k},
\eeq
and we will denote its complex conjugate by $E^*$. Here we have denoted $k_l = (\k \cdot \hat{e}_{l})$ and $\varepsilon_\iota(\k)_k = (\e_\iota(\k) \cdot \hat{e}_k)$.

In equation~\eqref{quad-hamiltonian}, $u^{I}$ and $X^{I}$ are the operator fields of the interaction picture, i.e. they evolve as free fields. As a reminder to the reader, the passage to the Heisenberg picture is implemented through
\beq
u (\x , t) = U^\dag (t,t_0)  u_I (\x , t)  U (t,t_0)
\eeq
with
\beq
U (t,t_0) = \text{T} \exp \left[ - i \int_{t_0}^{t} H_I (t') d t' \right]
\eeq
where $t_0$ is the time at which both operators coincide (typically it is taken to be $-\infty$).

One may ask what is the nature of the approximation involved in considering only the quadratic interaction terms for the quantum theory. Since the only way to be sure of the answer is to compute the sub-leading terms, we will leave this question unanswered until the last section, where we study explicitly the higher order corrections.

For the moment, we can start with the leading contribution to the phonon-phonon amplitude.

\subsection{Lowest order contribution in the phonon-string coupling}
\label{lowest_o}

Given that we are interested in phonon-phonon amplitudes, in order to have a non-vanishing contribution from~\eqref{ph-ph-a} we need to expand the time-ordered exponential up to quadratic order. Employing our definition for the string propagator and performing the relevant contractions, we see that
\begin{equation} \label{integrals2}
\begin{split}
\braket{f|i}|_{\mathcal{O}(b^2)} = - \frac{b^2 \mu^2 }{\hbar^2} & \int_{-\infty}^{\infty} \!\!\!\! dt \int_{-\infty}^{\infty} \!\!\!\!  dt'   \int_0^L  \!\!\!  ds \int_0^L \!\!\!  ds' \\ 
\times &  {\bf M}_{kl} {\bf M}_{k'l'} \Delta(s,s',t-t') \\  \times &  \braket{0 |a_{\iota}(\k^f) \frac{\partial u_k^I}{\partial x_l}(\x_0+(0,0,s),t) |0} \\ \times & \braket{0| \frac{\partial u_{k'}^I}{\partial x_{l'}}(\x_0+(0,0,s'),t')  a_{\iota'}^{\dagger}(\k^i) | 0}.
\end{split}
\end{equation}
If we define the $\mathcal T$ matrix through
\beq
\braket{f|i} \equiv 2\pi \, \delta(\omega_{\iota}(\k) - \omega_{\iota'}(\k')) \, \mathcal{T}, 
\eeq
then equation~\eqref{integrals2} yields, using (\ref{qdispl}),
\begin{equation}
\begin{split} \label{T-lowest}
\mathcal{T} = & \frac{2 i E^{*}(\k;\iota) E(\k'; \iota')}{ m L } \sum_{n=1}^{n_D}  \frac{(n\pi/L)^2}{\omega_n^2 - \omega^2} \\ & \,\,\,\,\,\,\,\,\,\,\,\,\,\,\,\,\,\, \times  \frac{e^{iL(\k' \cdot \hat{e}_3) - i n \pi} - 1}{(\k' \cdot \hat{e}_3)^2 - (n\pi/L)^2} \frac{e^{-iL(\k \cdot \hat{e}_3) + i n \pi} - 1}{(\k \cdot \hat{e}_3)^2 - (n\pi/L)^2}.
\end{split}
\end{equation}

An interesting feature of this result is that it is independent of $b$, as we have $m \propto b^2$; that is, the mass per unit length of the dislocation is proportional to the square of the Burgers vector. While it is too soon to make a definitive statement about the actual perturbative parameter of the problem, this suggests that $b$ \textit{per se} does not play that role, contrary to what one could at first think.

The observable that may be derived from the $\mathcal T$ matrix is the scattering cross-section. As we are interested in the outgoing flux, i.e. the angular distribution of phonons, given an incident phonon with well-defined momentum $\k'$, one obtains
\beq
\frac{d \sigma}{d \Omega} = \frac{\omega^2 |\mathcal{T}|^2}{4\pi^2 c_\iota^3 c_{\iota'}}.
\eeq
This dependence of the cross-section on the scattering amplitude holds independently of the order to which we compute observables in perturbation theory, which is exactly our next order of business.

\subsection{Phonon scattering by dislocation segments to all orders in the coupling}
\label{all_o}

Now we turn to solving for the scattering amplitude to all orders in the interaction Hamiltonian~\eqref{quad-hamiltonian}. That is, the computation of all terms in the power series development of the exponential that appears in (\ref{ph-ph-a}). The object of interest is the same as before~\eqref{ph-ph-a}, so we may proceed order by order and then sum the terms to reconstruct the result. It turns out that it will be sufficient to evaluate the fully connected parts of the amplitude, i.e. those which cannot be separated into independent factors, and that a power series results, that can be summed. Because the expressions involved in the derivation are quite lengthy, we shall present those details in Appendix~\ref{sec:details}, and here we will discuss the results.

In a very similar manner to the classical case \cite{Churochkin2016}, the $\T$ matrix one obtains is given by
\beq \label{exact-quadr}
\mathcal{T} =  {\v}^{\dagger}(\k) \cdot \frac{i  E^*(\k; \iota) E(\k'; \iota')}{1-{\bf T}(\omega)} \cdot {\v}(\k').
\eeq
The precise definitions of $\v$ and ${\bf T}(\omega)$ are listed in Appendix~\ref{sec:details}. For the purposes that with which we concern ourselves, $\v$ may be seen as an $n_D$-dimensional vector that contains information on how an incident wave interacts with the each mode of the string, to first order. This is readily seen by noticing that the amplitude to lowest order~\eqref{T-lowest} is given by $i E^*(\k, \iota) E(\k',\iota') {\v}^{\dagger}(\k) \cdot {\v}(\k')$. 

On the other hand, ${\bf T}(\omega)$ may be regarded as an $n_D \times n_D$ symmetric matrix quantifying the internal structure of the theory, being the object wherein to look for a perturbative parameter. This object has a rich structure, which will study thoroughly in the following subsections.

\subsubsection{Finiteness of the theory}

In previous works studying the problem of phonon scattering by dislocations \cite{Churochkin2016}, a renormalization procedure was necessary to guarantee a finite result at every order in perturbation theory. Specifically,
one may be worried that the integral (\ref{mixed-prop-a}) involved in the definition of ${\bf T}$ may diverge:
\begin{equation} \label{mixed-prop}
\begin{split}
F(n,n') \equiv & \frac{\mu b {\bf M}_{kl} }{ \sqrt{\rho}} \int \frac{d^3k}{(2\pi)^3} k_l k_{l'} \int_0^L \!\!\! ds  \sin\left(\frac{n \pi s}{L} \right) e^{i k_3 s} \\ 
& \times  \left[ \frac{\delta_{kk'} - \hat{k}_k \hat{k}_{k'}}{\omega_T(\k)^2 - \omega^2 - i\epsilon} +  \frac{\hat{k}_k \hat{k}_{k'}}{ \omega_L(\k)^2 - \omega^2 - i\epsilon } \right] \\ 
& \times  \int_0^L \!\!\!  ds' \sin\left(\frac{n' \pi s'}{L} \right) e^{-i k_3 s'} \frac{\mu b {\bf M}_{k'l'}}{\sqrt{\rho}},
\end{split}
\end{equation}
where $\epsilon$ is a positive infinitesimal and $k_i \equiv \k \cdot \hat{e}_i$. This object may be regarded as a correction to the propagator of the string modes, as it only involves indices labelling these states.

It turns out that one can perform the integration over $|\k|$ by employing the prescription defined by $\epsilon$. After doing some residue calculus and performing the integral over the azimuthal angle, one finds
\begin{equation} \label{finite-result}
\begin{split}
F(n,n') =& \, i \omega^3 \frac{ \mu^2 b^2 }{8 \pi \rho } \int_0^L \!\!\!  ds \int_0^s \!\! ds' \! \int_0^1 \!\!  du \, (1 -u^2)   \\ & \times \left[\sin\left(\frac{n \pi s}{L} \right) \sin\left(\frac{n' \pi s'}{L} \right) + (n \leftrightarrow n') \right]  \\ & \times \left[ \frac{1+u^2}{c_T^5} e^{i(s-s')\omega u/c_T } + \frac{1-u^2}{c_L^5} e^{i(s-s')\omega u/c_L } \right].
\end{split}
\end{equation}
As expected \cite{Churochkin2016}, the real part of~\eqref{finite-result} vanishes as $\omega \to 0$ faster than the imaginary part, which goes as $\omega^3$ at low frequencies. Furthermore, it is manifestly finite at all frequencies, as every piece of the integrand is bounded.

\subsubsection{Low and high frequency behavior}

It is instructive to explore the behavior of the preceding formulae in the low-frequency limit $ \omega L/c \propto kL \ll 1$ to see whether this result is consistent with previous findings. As $W_{\iota}$ is finite and nonzero for small $\omega$, we may simply evaluate it at $\omega=0$ and only keep the leading factor $\omega^3$. To further simplify matters, we will only take into account the dominant contribution to scattering at large wavelengths: $n = n' = 1$. We then obtain
\beq \label{lowfreqT}
{\bf T}_{\omega \ll c_T/L} \simeq i \omega^3 \frac{\mu^2 b^2 L}{m \rho \omega_1^2 c_T^5} \frac{4}{5\pi^3} \left[1 + \frac{2}{3\gamma^5} \right]
\eeq
where we have also approximated $\omega_1^2 - \omega^2 \approx \omega_1^2$ in the denominator. This is exactly what was inferred in previous papers\cite{Churochkin2016} for the value of this operator, thus verifying that ${\bf T}$ is purely imaginary at low frequencies.

Now we may infer the perturbative parameter involved in expression~\eqref{exact-quadr} at low frequencies. As it turns out, equation~\eqref{lowfreqT} is proportional to $(\omega L/c_T)^3$. Although it might not be the fundamental constant we could have hoped for (analogous to the fine structure constant), it is physically sound: the larger the characteristic wavelength of a wave packet, the less it is affected by dislocations.

On the other hand, in the high-frequency limit the integrals in~\eqref{finite-result} vanish as the integrand turns highly oscillatory. This result is generically known as the \textit{Riemann-Lebesgue lemma}. If one performs a change of variables $s = xL$ and then takes $L \to \infty$, one concludes that the integrals should vanish on the same grounds. However, this assumes that we leave $n,n'$ as constants, which is not of much physical interest. If we instead take $k_z \equiv n\pi/L$ fixed when $L$ goes to infinity, we recover a continuum of modes in an infinitely long string, and as the system gains a translation symmetry along the $z$ axis, wavenumber conservation along this direction emerges in the scattering amplitudes. Mathematically, this feature arises, again, due to the \textit{Riemann-Lebesgue} lemma, as the integrand becomes highly oscillatory unless the wavenumbers $\omega u/c_\iota$ are equal. A detailed study of this regime will be undertaken in future work.

\subsubsection{The intermediate region: a numerical analysis}

To study the behavior of the amplitude~\eqref{exact-quadr}, or equivalently the cross-section, in the regime where the frequency $\omega$ is close to the resonant frequencies of the string we will resort to numerical computations. For these purposes, we find more instructive to define a dimensionless frequency $\nu \equiv \omega L/c_T$ and write $\mathcal{T}$ as
 \beq
 \begin{split}
 \T =& \, \frac{ \rho b^2 L^2 c_T^3}{m  c_\iota c_{\iota'}}  {\bf w}^{\dagger}(\k) \cdot \frac{i \nu}{1-{\bf T}(\omega)} \cdot {\bf w}(\k') \\ & \times \hat{k}_l {\bf M}_{kl} \varepsilon_\iota(\k)_k  \hat{k'}_{l'} {\bf M}_{k'l'} \varepsilon_{\iota'}(\k')_{k'} e^{i(\k' -\k) \cdot \x_0}
 \end{split}
 \eeq
 where ${\bf w}$ is a $n_D$-dimensional vector with entries
 \beq
 [{\bf w}(\k)]_n \equiv \frac{n\pi}{\frac{L}{c_T} \sqrt{\omega_n^2 - \omega^2} } \frac{e^{iL(\k \cdot \hat{e}_3) + i n \pi} - 1}{(L\k \cdot \hat{e}_3)^2 - (n\pi)^2}.
 \eeq
Thus, we see that the usual scattering amplitude $f$ such that $d\sigma/d\Omega = |f|^2$ is given by
\beq \label{scatt-usual}
\begin{split}
f(\omega, \k, \k') \equiv& \, \frac{ \rho b^2 L c_T^4}{2\pi m   \sqrt{c_\iota^5 c_{\iota'}^3}} {\bf w}^{\dagger}(\k) \cdot \frac{i \nu^2}{1-{\bf T}(\omega)} \cdot {\bf w}(\k') \\ & \times \hat{k}_l {\bf M}_{kl} \varepsilon_\iota^*(\k)_k  \hat{k'}_{l'} {\bf M}_{k'l'} \varepsilon_{\iota'}(\k')_{k'} e^{i(\k' -\k) \cdot \x_0},
\end{split}
\eeq
where the vectors $\k,\k'$ satisfy $\omega = |\k| c_{\iota} = |\k'| c_{\iota'}$.

\begin{figure*}[t!]
\includegraphics[scale=0.4]{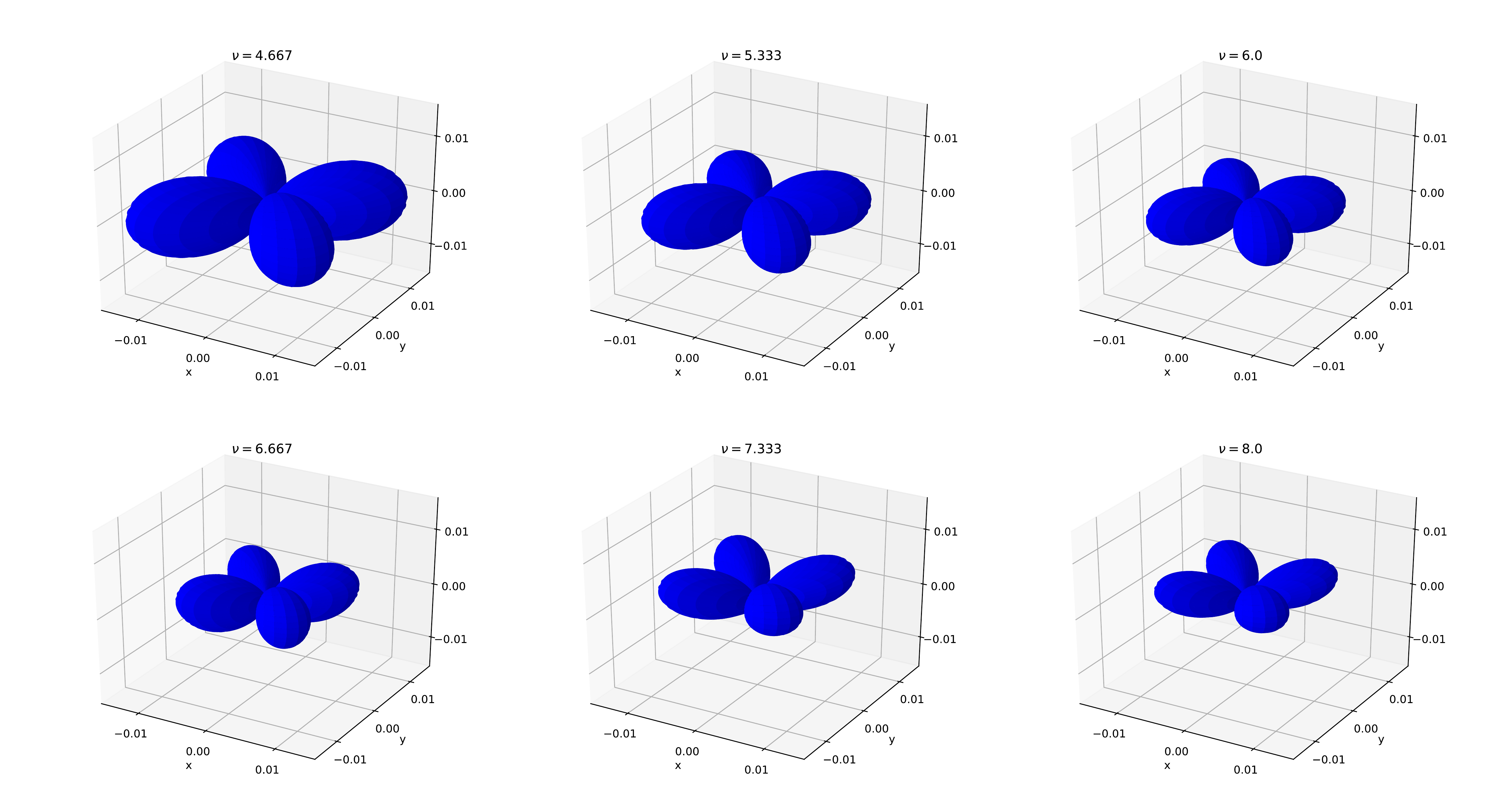} 
\caption{Absolute value of the scattering amplitude $|f|$ for phonons by dislocation segments for $n_D = 10$, $\gamma=2$, $\ln (\delta/\delta_0) = 3$, with both incident and outgoing phonons having longitudinal polarization, as a function of the outgoing direction $\hat{k}$. The ingoing direction was set to $\hat{k}' = (1/2,1/2,1/\sqrt{2})$. Different values for the incident frequency are shown in different plots. The coordinates $(x,y)$ correspond to the directions $(\hat{e}_1, \hat{e}_2)$ respectively. The plots are to scale with respect to each other. Note that at low frequencies there is an up-down symmetry that is lost when $\nu$ is gradually increased above $\approx 2\pi$, that is, at frequencies higher than the second string resonance, which is the first that can distinguish $\hat{e}_3$ from $-\hat{e}_3$ and is thus sensitive to the direction of incidence $\hat{k}'$.}
\label{fig:scatt-amp}
\end{figure*}

To show an example out of this, Figure~\ref{fig:scatt-amp} shows the absolute value of the longitudinal to longitudinal scattering amplitude for a range of frequencies in a typical scattering event. From low frequencies until the first resonance $\omega_1$ the angular distribution is concentrated around the $\hat{e}_1 - \hat{e}_2$ plane. However, after the first resonance the cross-section starts experiencing the effects of  the incident direction, obtaining asymmetric patterns with respect to $\hat{e}_3$. Similar results can be obtained for cross-sections involving transverse polarization; the main difference is that the angular distribution changes because of the different polarization directions of the incident wave.

Having the differential scattering cross section for the scattering of phonons by dislocation segments, the next step is to compute the total cross section for a given mode of transmission. This is obtained by integrating over all outgoing modes of polarization and momenta consistent with the kinematical restrictions. In this case, these restriction amount to energy conservation. However, the resulting cross-section depends on the incident distribution of polarization, which is particularly relevant in the case of transverse polarization. In a general situation, given an initial density matrix $\hat{\rho}_{\k,T}^0$ for the transverse modes of wavenumber $\k$, we can compute the fraction of scattered phonons as
\beq
{\rm tr} (U \hat{\rho}_{\k,T}^0 U^{\dagger} P_{\{\k' : \k' \neq \k\}} )
\eeq
where $P_{\{\k' : \k' \neq \k\}}$ is an orthogonal projector to the space of all single-particle states with wavenumber different than $\k$, which plays the role of selecting the scattered piece of the time-evolved state.

\begin{figure*}[t!]
\includegraphics[scale=0.55]{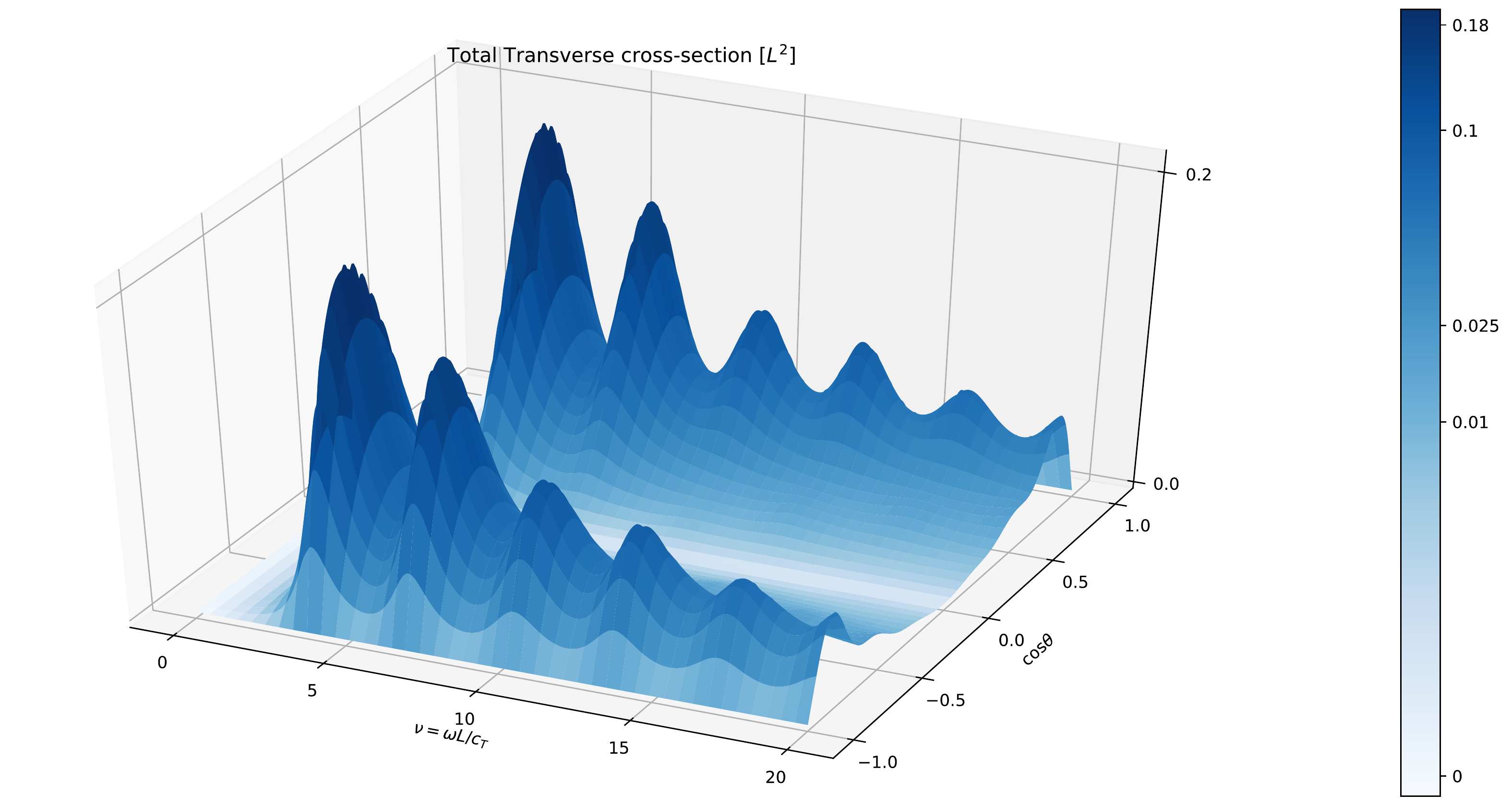} 
\includegraphics[scale=0.5735]{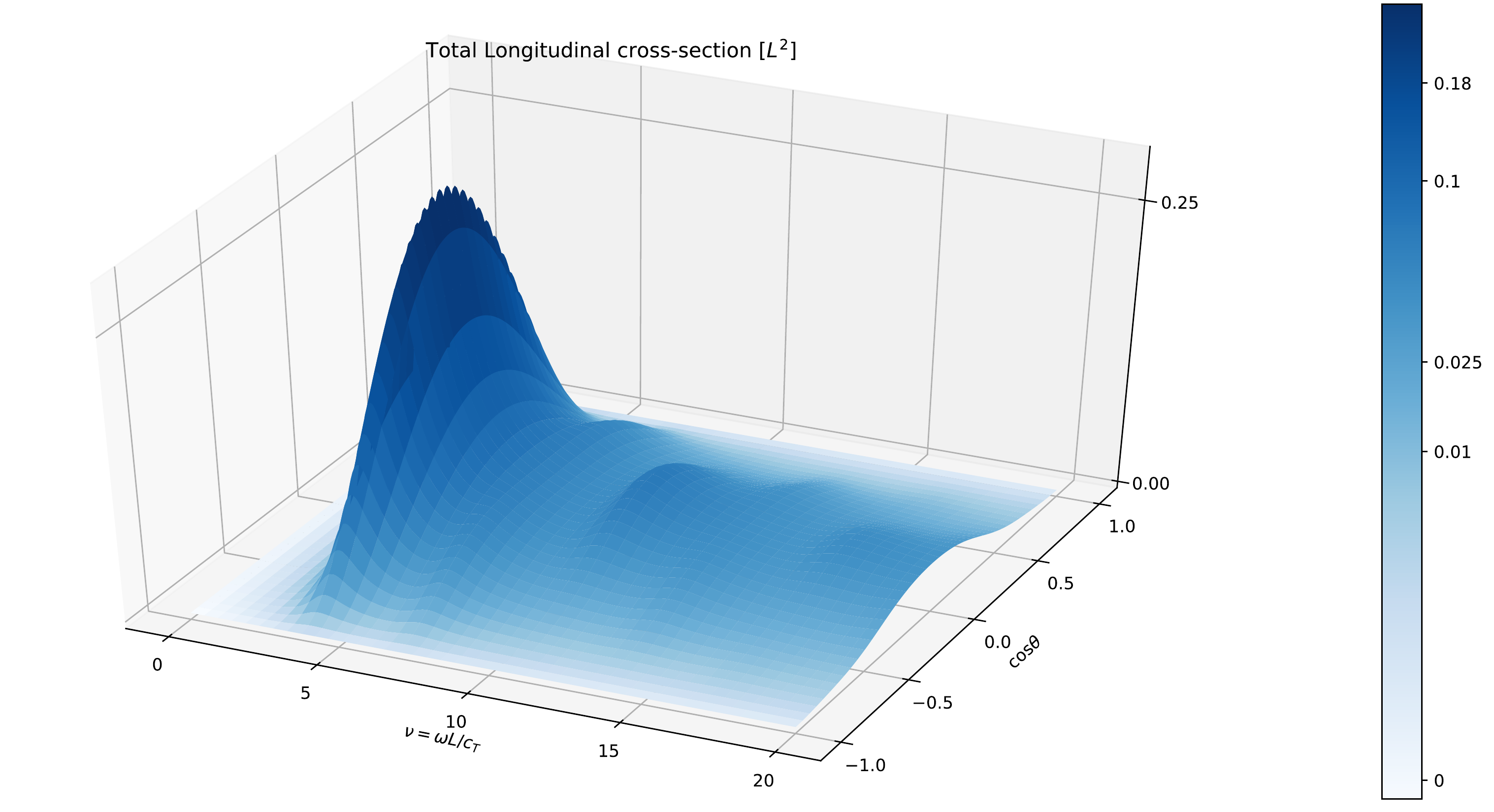} 
\caption{Surface plot of the scattering cross-sections $\sigma_T$ and $\sigma_L$ in log-linear scale, as a function of frequency and angle of incidence, for phonons by oscillating dislocation segments of finite length. In the figure, $n_D = 10$, $\gamma=2$, and $\ln (\delta/\delta_0) = 3$. Furthermore, we have defined $\cos \theta = \hat{k} \cdot \hat{e}_3$ and chosen $ |\hat{k} \cdot \hat{e}_1| =|\hat{k} \cdot \hat{e}_2|$ as a representative instance of these quantities. Note the rich behavior of the cross sections as a function of frequency, in contrast with the linear-in-frequency behavior inferred from the scattering by an infinitely long, static, dislocation\cite{Klemens1955,Carruthers1959}.}
\label{fig:scatt-cross}
\end{figure*}

Dividing the result by the incident flux, it is not difficult to see that for an unpolarized incident beam of transverse-polarized phonons, the total cross-section is given by
\beq
\begin{split}
\sigma_T(\k) =& \, \frac{\pi L^2}{2} \left(\frac{ \rho b^2}{2 \pi m} \right)^2 \left( |\hat{k} \times \hat{e}_3|^2 - 4(\hat{k} \cdot \hat{e}_1)^2 (\hat{k} \cdot \hat{e}_2)^2  \right) \\ \times & \!  \int_{-1}^1 \!\!\!\! du \! \left[ (1 -u^4) \left| {\bf w}^{\dagger}_T(u,\nu) \frac{\nu^2}{1 - {\bf T}(\nu) } {\bf w}_T(\hat{k} \cdot \hat{e}_3,\nu)  \right|^2 \right. \\ & \left. + \frac{(1-u^2)^2}{\gamma^5} \left| {\bf w}^{\dagger}_L(u,\nu) \frac{\nu^2}{1 - {\bf T}(\nu) } {\bf w}_T(\hat{k} \cdot \hat{e}_3,\nu)  \right|^2 \right]
\end{split}
\eeq
where $\{\hat{e}_i\}_i$ is the orthonormal basis used to describe the string, and we have defined
\bea
\!\!\!\!\!\!\!\!\!\!\!\! [{\bf w}_T(u,\nu)]_n &\equiv& \frac{n\pi}{\frac{L}{c_T} \sqrt{\omega_n^2 - \nu^2 c_T^2/L^2} } \frac{e^{i u \nu + i n \pi} - 1}{u^2 \nu^2 - (n\pi)^2}, \\
\!\!\!\!\!\!\!\!\!\!\!\! {[}{\bf w}_L(u,\nu)]_n &\equiv& \frac{n\pi}{\frac{L}{c_T} \sqrt{\omega_n^2 - \nu^2 c_T^2/L^2} } \frac{e^{i u \nu/\gamma + i n \pi} - 1}{u^2 \nu^2/\gamma^2 - (n\pi)^2}.
\eea
as a function of the dimensionless frequency $\nu$ and the angle of the corresponding wavenumber $\k'$ with respect to the axis of the string $u = \hat{k}' \cdot \hat{e}_3$. Similarly, in the case of longitudinal polarization we have
\beq
\begin{split}
\sigma_L(\k) =& \, \frac{\pi L^2}{\gamma^3} \left(\frac{ \rho b^2}{2 \pi m} \right)^2 \left( 4(\hat{k} \cdot \hat{e}_1)^2 (\hat{k} \cdot \hat{e}_2)^2  \right) \\ \times  & \!  \int_{-1}^1 \!\!\!\! du \! \left[ (1 -u^4) \left| {\bf w}^{\dagger}_T(u,\nu) \frac{\nu^2}{1 - {\bf T}(\nu) } {\bf w}_L(\hat{k} \cdot \hat{e}_3,\nu)  \right|^2 \right. \\ & \left. + \frac{(1-u^2)^2}{\gamma^5} \left| {\bf w}^{\dagger}_L(u,\nu) \frac{\nu^2}{1 - {\bf T}(\nu) } {\bf w}_L(\hat{k} \cdot \hat{e}_3,\nu)  \right|^2 \right].
\end{split}
\eeq
Plots of these results are shown in Figure~\ref{fig:scatt-cross}.

It is important to stress that the cross-section for transverse polarization is dependent on the particular polarization of the incident wave, as both the polarization vector $\e$ and the wavenumber $\k$ must have projections over $\hat{e}_1$ or $\hat{e}_2$ to get a nonzero result. This can be readily appreciated by looking at the scattering amplitude~\eqref{scatt-usual}. For now, and for all forthcoming purposes, we will proceed using the average cross-section for transverse modes, which is exactly what one obtains with circular polarization, or with unpolarized phonons.

As a final note on this section, let us comment on the role of the cutoff $n_D$. Typically, it should be chosen so that the wavenumber $n\pi/L$ matches the Debye wavelength. However, if one is interested in making predictions at frequencies below a given eigenfrequency of the string $\omega_{n^*}$, it is possible to choose the cutoff as $n^*$ without losing predictive power. In this sense, the results shown in the plots of Figures~\ref{fig:scatt-amp} and~\ref{fig:scatt-cross} always hold (independently of our choice $n_D = 10$), provided that, roughly speaking, $\nu < 30$.

\subsubsection{Resonances and decay rates}

A recurring feature within the expressions for the phonon-phonon scattering amplitude we have written down so far is the apparent existence of poles at specific frequencies, namely at the natural frequencies of oscillation of the string. However, none of these creates any singularity, as should be obvious from Figure~\ref{fig:scatt-cross}: the width and height of the peaks in the cross section is finite. 

This is related to the fact that phonons at these frequencies can decay into excited string states, thus aborting the resonant behavior of the interaction and giving the peak a finite width.

Indeed, the interaction between the string and a phonon with coinciding eigenfrequency will generate asymptotic states (i.e. those that result from a scattering process) that assign some probability to having an excited string. Conversely, an initially excited string is also an unstable particle, and it will have a decay rate by emission of phonons, with a probability amplitude equal to the phonon-string decay up to an overall sign.

For instance, the decay rate of an excited string may be evaluated by the same means that we evaluated the phonon-phonon cross-sections. The object of interest is
\beq 
\langle f|i\rangle = \braket{0| a_{\iota}(\k) \text{T} \exp \left[ - \frac{i}{\hbar} \int_{-\infty}^{\infty} H_I(t) dt \right] a_n^{\dagger}  |0},
\eeq
which after a completely analogous process, in terms of the $\T$ matrix $\langle f|i\rangle = (2\pi) \delta(\omega_\iota - \omega_n) \T$, gives
\beq
\label{sponT}
\T = {\bf v}^{\dagger}(\k) \cdot \frac{- i E^*(\k;\iota) G_n}{1 - {\bf T}(\omega)} \cdot {\bf e}_n,
\eeq
where we have defined $G_n \equiv n\pi/\sqrt{mL^3 \omega_n}$ and ${\bf e}_n$ as a unit vector in the $n_D$-dimensional space where $\bf T$ acts upon, with a nonzero component for the $n$-th mode only. The amplitude of the inverse process (phonon to string) is given by minus the complex conjugate of this expression.

Finally, the rate of spontaneous emission for excitations of the string in its mode $n$ is given by
\beq
\Gamma_n = \frac{\omega_n^2}{4\pi^2} \sum_{\iota} \frac{1}{c_\iota^3} \int d\Omega |\T|^2,
\eeq
with ${\T}$ given by (\ref{sponT}), which gives the typical lifetime of an excited string.

The converse process, i.e. the absorption of a phonon by a dislocation segment, is also allowed, and as such it contributes to the total cross-section of a phonon traveling through the solid. 
However, since this scattering process will be subject to energy conservation, the corresponding amplitude will be either zero or formally infinite via the presence of the Dirac delta. Therefore, it will not contribute to the bulk thermal conductivity: if the amplitude is zero, the total cross-section will be described purely by phonon-phonon processes, and if it is infinite, then that specific mode channel will contribute zero, which renders it irrelevant as we have a continuum of modes of transmission available. Although, the situation should change in the presence of a continuous distribution of lengths, because in this situation absorption to a continuum of modes will be possible, and thus the macroscopic observables, such as the thermal conductivity, should be modified.

\subsection{Dislocations in a mesoscopic medium}
\label{mass_op}

In this section we discuss some methods and approximations that should be useful for dealing with a dislocation-filled medium from a mesoscopic perspective. Firstly we will write down an expression useful in situations where the scattering is well approximated by a single interaction with a dislocation, imposing that these interactions should, on average, behave as if the medium were homogeneous and isotropic. We also present a compact expression for the propagator in this approximation. Then we proceed to outline the general situation, where this approximation may or may not hold true.

\subsubsection{An effective mass operator for weak and independent scattering}

In many realistic situations, we will have randomly distributed dislocations within the solid. Therefore, if we want to make practical predictions on phonon scattering within a solid, we must account for this in some way. The \textit{Weak and Independent Scattering Approximation} \cite{vanRossum} provides a solution to this problem: average the $\T$ matrix over the internal degrees of freedom of the system. In this case, we will average over the dislocation position $\x_0$ and orientation $(\hat{e}_1, \hat{e}_2, \hat{e}_3)$. For computational simplicity, however, we will keep a unique string length.

It should be clear from equation~\eqref{T-lowest} that averaging over the position $\x_0$ of the string will give a Dirac delta in momentum space, thus enforcing momentum conservation by demanding the sample to have a homogeneous distribution of dislocations. The result of averaging over dislocation directions should then yield an isotropic result, and since $\delta_{ij}$ and $ \hat{k}_i \hat{k}_j$ are the only rank-2 tensors available,  it should be possible to express the average $\T$ matrix $\bar{\T}$ as
\beq
\begin{split}
\bar{\T} = & \, i(2\pi)^3 \delta^{(3)}(\k - \k') \varepsilon_{\iota}(\k)_i  \varepsilon_{\iota'}(\k')_j  \frac{n_d L \mu^2 b^2 k^2}{8 m \rho \omega}  \\
& \times \left( (\delta_{ij} - \hat{k}_i \hat{k}_j) \, {\rm tr}_{n_D} \!\!\left[ \frac{{\bf F}_T(k,\omega)}{1-{\bf T}(\omega)} \right] \right. \\ & \left. \,\,\,\, \,\,\,\, \,\,\,\, \,\,\,\, \,\,\,\, \,\,\,\,\, + \hat{k}_i \hat{k}_j \, {\rm tr}_{n_D} \!\! \left[ \frac{{\bf F}_L(k,\omega)}{1-{\bf T}(\omega)}\right]  \right),
\end{split}
\eeq
where $n_d$ is the dislocation number density and ${\rm tr}_{n_D}$ stands for taking the trace over the different modes of excitation of the string.
Once the angular average is performed, one obtains
\beq
\begin{split}
[{\bf F}_T]_{n'n} = & \, \frac{n n' \pi^2}{\sqrt{\omega_n^2 - \omega^2} \sqrt{\omega_{n'}^2 - \omega^2 }} \int_{-1}^1 \!\!\! du (1- u^4)  \\ & \times   \frac{e^{ikLu}(-1)^{n'} - 1}{(k L u)^2 - (n'\pi)^2}  \frac{e^{-ikLu} (-1)^n - 1}{(k L u)^2 - (n\pi)^2}
\end{split}
\eeq
and
\beq
\begin{split}
[{\bf F}_L]_{n'n} = & \, \frac{n n' \pi^2}{\sqrt{\omega_n^2 - \omega^2} \sqrt{\omega_{n'}^2 - \omega^2 }} \int_{-1}^1 \!\!\! du \, 2 (1- u^2)^2  \\ & \times   \frac{e^{ikLu}(-1)^{n'} - 1}{(k L u)^2 - (n'\pi)^2}  \frac{e^{-ikLu} (-1)^n - 1}{(k L u)^2 - (n\pi)^2}.
\end{split}
\eeq
The details of these computations are presented in Appendix~\ref{sec:angular}. These results are consistent with those of previous works \cite{Maurel2005a, Maurel2005b, Churochkin2016} in the large wavelength limit $kL \ll 1$, and furthermore completely define the scattering amplitude. 

Now we may derive a macroscopic mass operator for the phonon dispersion relation from this effective scattering amplitude. Usually one defines the mass operator $\Sigma$ as the operator that solves the Dyson equation:
\beq
\langle G \rangle = [(G^0)^{-1} - \Sigma]^{-1}
\eeq
where $G^0$ is the Green's function for the elastic continuum without dislocations, which (up to normalization) we write as
\beq
(G^0)^{-1}_{ij} =  - \omega^2 \delta_{ij} + c_T^2 k^2 (\delta_{ij} - \hat{k}_i \hat{k}_j)  + c_L^2 k^2  \hat{k}_i \hat{k}_j ,
\eeq
and $\langle G \rangle$ is an effective Green's function, resulting from performing the averages just described. By inspecting the time-ordered two-point function associated to the amplitude~\eqref{ph-ph-a}. If we define a reduced amplitude  $\bar{\T}_{ij}$ through 
\beq
\bar{\T} = i(2\pi)^3 \delta^{(3)}(\k - \k') \frac{\varepsilon_{\iota}(\k)_i  \varepsilon_{\iota'}(\k')_j}{2 \omega} \bar{\T}_{ij}
\eeq
then the effective Green's function is given by
\beq
\langle G \rangle_{ij} = G^0_{ij} + G^0_{ik} \bar{\T}_{kl} G^0_{lj} 
\eeq
However, this considers single scattering through a single mesoscopic region in space. If we take this, as we stated earlier, to represent a single (effective) scattering process, this would just be the first term for a macroscopic Born series, which would read
\bea
\langle G \rangle_{ij} &=& G^0_{ik} \sum_{n = 0}^{\infty} [(\bar{\T} G^0 )^n]_{kj} \nonumber \\
&=& ((G^0)^{-1} - \bar{\T} )^{-1},
\eea
from where we get the macroscopic mass operator as
\bea
\Sigma_{ij} &=&  \bar{\T}_{ij} \nonumber \\
&=&  \frac{n_d L \mu^2 b^2 k^2}{4 m \rho} \left( (\delta_{ij} - \hat{k}_i \hat{k}_j) \, {\rm tr}_{n_D} \!\!\left[ \frac{{\bf F}_T(k,\omega)}{1-{\bf T}(\omega)} \right] \right. \nonumber \\ & & \left. \,\,\,\, \,\,\,\, \,\,\,\, \,\,\,\, \,\,\,\, \,\,\,\, \,\,\,\, \,\,\,\, \,\,\,\, \,\,\,\, + \, \hat{k}_i \hat{k}_j \, {\rm tr}_{n_D} \!\! \left[ \frac{{\bf F}_L(k,\omega)}{1-{\bf T}(\omega)}\right]  \right),
\eea
which represents, as the name suggests, a frequency-dependent mass term to be added to the Green's function. It should be noted that it also includes dissipative terms coming from the interaction with the strings.

In order to illustrate this result, and given that the spirit of the approximation is to model an elastic medium at a mesoscopic scale, let us take $n_D = 1$. This is justified as long as the elastic waves of interest have wavelengths larger than $L$, or equivalently frequencies below $c_T/L$. 

The dispersion relation that determines how elastic waves propagate is given by the poles of the effective Green's function $\langle G \rangle$. Because ${\bf F}_T$, ${\bf F}_L$, and ${\bf T}$ are generically complex-valued, one can expect dissipative effects to emerge. Moreover, by simple inspection one can determine that the presence of dislocations changes the sound speed at arbitrarily low wavenumber/frequency. The effective sound speeds at very long wavelengths $kL \ll 1$ are given by
\bea
c_{T, {\rm eff}}^2 &=& c_T^2 \left( 1 - \frac{8}{5\pi^2} \frac{n_d L c_T^2}{\omega_1^2} \frac{\rho b^2}{m} \right), \\
c_{L, {\rm eff}}^2 &=& c_L^2 \left( 1 - \frac{32}{15\pi^2} \frac{n_d L c_L^2}{\gamma^4 \omega_1^2} \frac{\rho b^2}{m} \right),
\eea
which are exactly those found with the classical theory in previous works~\cite{Maurel2005b} in the $\omega \to 0$ limit.

\begin{figure}[t!]
\includegraphics[scale=0.55]{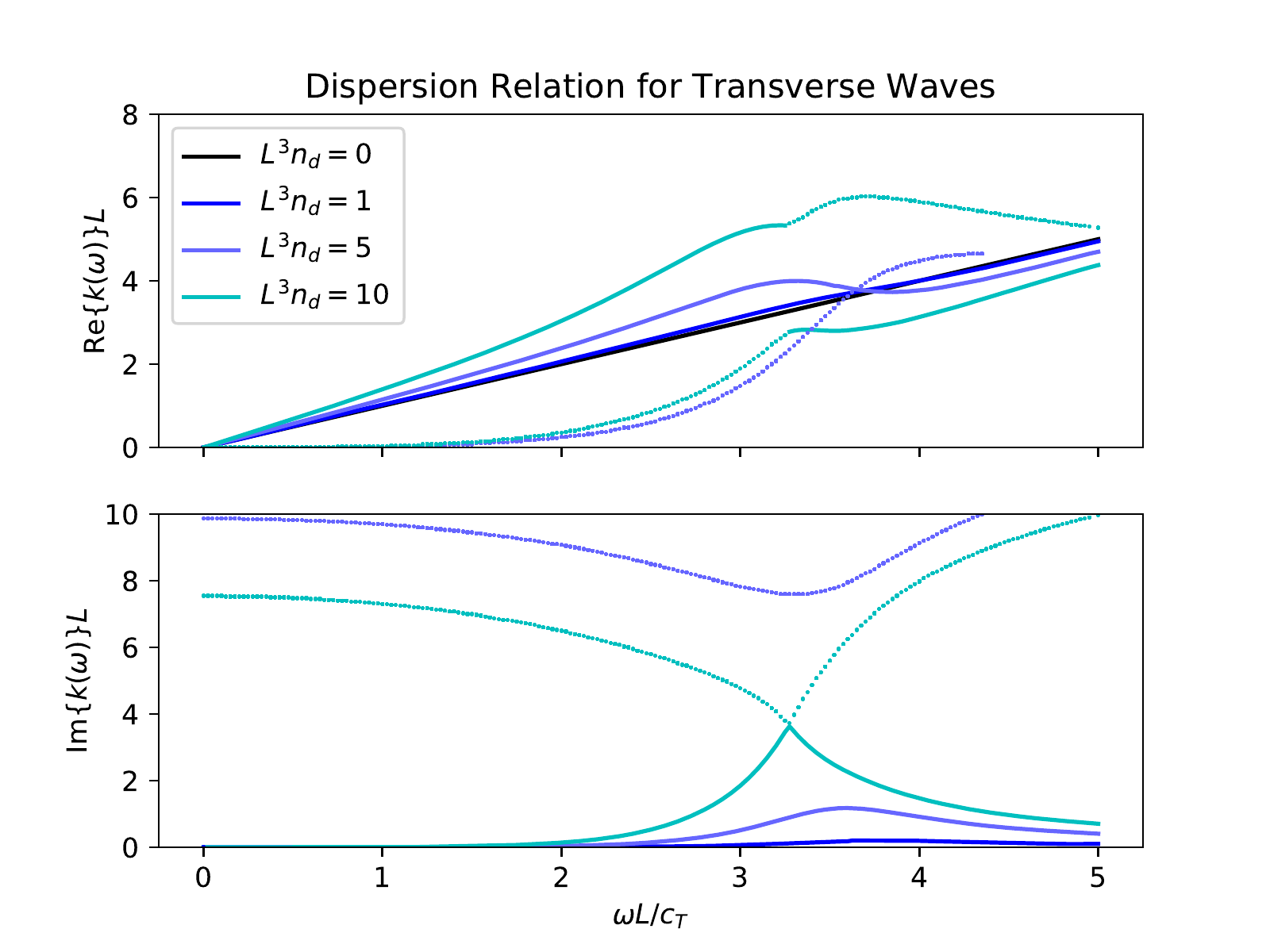} 
\includegraphics[scale=0.55]{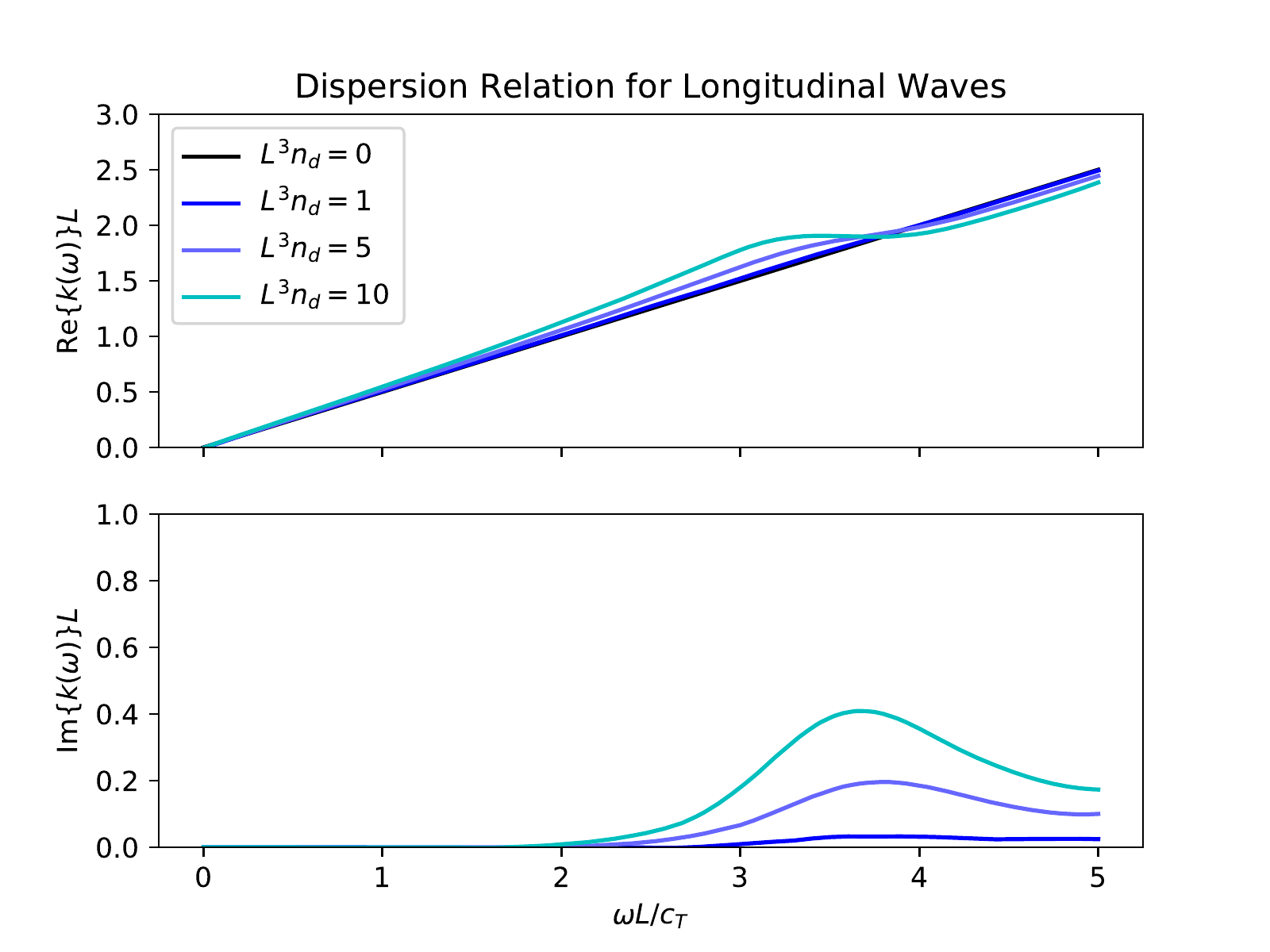} 
\caption{Numerical solution of the dispersion relation $k(\omega)$ for the propagation of phonons through a medium with many, randomly located and oriented, dislocation segments for different values of the dislocation density $n_D$. In the plots, we have set $\gamma=2$ and $\ln (\delta/\delta_0) = 3$. Note that shear phonons are more strongly affected that lomgitudinal phonons. The propagative solutions are plotted with continuous lines and the evanescent waves with discontinuous lines. See text for a discussion.}
\label{fig:disper-rel}
\end{figure}

To find the full dispersion relation, we solve for the location of the poles numerically, and present the results in Figure~\ref{fig:disper-rel}. These were computed searching for the propagative solution of the poles, i.e., that with the lowest value for ${\rm Im} (k(\omega) )$. The curves shown have been smoothed by a Savitzky-Golay filter. In the case of transverse wave propagation, we also present, with discontinuous lines, the dispersion relation for evanescent waves that appear because of having a complex dispersion relation. 

We start discussing the propagative solutions. Qualitatively speaking, at frequencies below the first resonance the sound speed of the medium decreases and the medium gets increasingly dissipative as the frequency is increased. A maximum is reached in the imaginary part of the wavenumber at a frequency close to that of the first resonance peak of the scattering amplitude, at which point the real part of the wavenumber crosses the line of the free theory dispersion relation, and then starts moving away from the resonant behavior, re-approaching the free theory behavior at larger frequencies. However, if the frequency is further increased, one should expect the effects of the next resonance peaks to kick in.

At first sight in Figure~\ref{fig:disper-rel}, the group velocity $\partial \omega/ \partial k$ appears to exceed the free theory sound speed at frequencies close to the resonance peak. However, as J. D. Jackson points out in his classic textbook~\cite{Jackson:1998nia}, the notion of group velocity through $\partial \omega/ \partial k$ is well-defined when the variation of the refraction index $n(\omega)$ is small, a condition that is utterly demolished in regions of anomalous dispersion, which is exactly the case at hand. Therefore, one should not worry  about the apparently causality-violating behavior of ${\rm Re}(k(\omega))$ that would appear around the resonance peak if one took the group velocity definition as meaningful for all frequencies.

Finally, we point out that the presence of evanescent waves for transverse polarization and not for longitudinal polarization in the range of frequencies/wavenumbers explored in the plots is mostly due to the dislocations having combinations of eigenfrequencies/eigenstates that resemble more closely to $\omega = kc_T$ than to $\omega = kc_L$. This is because in $\omega_n = \frac{n \pi}{L} c_{d} $, we can identify $\frac{n \pi}{L}$ as the wavenumber, and $c_d $ will always satisfy $c_d < \sqrt{2} c_T$ (in the plot, $c_d \approx 1.18 c_T$). Therefore, for longitudinal waves it is less common to satisfy $\omega \sim \omega_n$ and $\k \cdot \hat{e}_3 \sim \frac{n\pi}{L}$ simultaneously than for transverse waves. Thus, as longitudinal waves will consequently have fewer interactions with the dislocations, their dispersion relation is less affected. Nonetheless, evanescent solutions should also exist, albeit with a shorter decay length. In this sense, increasing the dislocation density makes it possible to have states that are more extended throughout the solid, as one might have intuited from the beginning.

Further analysis of the critical dislocation density $n_d L^3$ at which the propagating waves' dispersion relation intersects with that of evanescent waves remains pending for future projects. It would be of interest to explore links, if any, and possible research directions, with metamaterials, by trying to tune the mass operator we have just derived to serve a specific purpose. On the other hand, it is possible that in such contexts, or in the present context, the Weak and Independent scattering starts to deviate from the physical phenomena, precisely because of a high dislocation density that makes independent scattering unfeasible. 

Therefore, we feel compelled to at least outline what a more general situation would be, and how the quantities that previously led to the mass operator would be computed.

\subsubsection{Many dislocations: the general case}

Formally, we may write the interaction Hamiltonian for a medium with many dislocations as
\beq
\begin{split}
H_I = & \, \int_{\x_0, L, b, \sphericalangle} \!\!\!\! \!\!\!\! \!\!\!\! \!\!\!\!  f(\x_0, L, b, \sphericalangle) \int_0^L \!\!\! ds \,  \mu b {\bf M}_{kl}(\sphericalangle) \\ & \times \frac{\partial u^{I}_k}{\partial x_l}(\x_0+s \hat{e}_3(\sphericalangle) ,t) X^{I}(s,t;\x_0, L, b, \sphericalangle)
\end{split}
\eeq
where $f(\x_0, L, b, \sphericalangle)$ is some distribution of positions $\x_0$, lengths $L$, Burgers vector magnitude $b$, and orientation $\sphericalangle$ characterizing the triad defined by the equilibrium position of the string and the direction of the Burgers vector $\hat{b}$.
This translates into new commutation relations for the mode operators of the string, which now are also a function of $\x_0, L, b, \sphericalangle$:
\beq
\begin{split}
& [a_n(\x_0, L, b, \sphericalangle), a_m^{\dagger}(\x_0', L', b', \sphericalangle')] \\ &= \delta_{nm} \delta^{(3)}(\x_0 - \x_0') \delta(L - L') \delta(b - b') \delta( \sphericalangle - \sphericalangle' )
\end{split}
\eeq
with the other commutators vanishing.

In this setting, we now have a dislon field $X(s,t;\x_0, L, b, \sphericalangle)$ over the parameters of the string, with $f(\x_0, L, b, \sphericalangle)$ describing how the field is arranged and how the positions $\x_0$ are correlated with the other variables. 

Even though this may seem to be significantly more complicated than what we have previously studied, some generalities are at hand. For instance, it is fairly easy to see that now the effective propagator $F(n,n')$ is now promoted to a function of not only the mode of each interacting string, but also of their position, orientation, and characteristic lengths. Concretely, we have
\begin{equation} 
\begin{split}
F =& \, i \omega^3 \frac{ \mu^2 b b' {\bf M}_{ik}(\sphericalangle) {\bf M}_{jl}(\sphericalangle') }{4 \pi^2 \rho } f(\x_0, L, b, \sphericalangle) f(\x_0', L', b', \sphericalangle')  \\ & \times \int d\Omega_+ \, \hat{k}_k \hat{k}_l   \int_0^L \!\!\!  ds \int_0^{L'} \!\!\! ds' \sin\left(\frac{n \pi s}{L} \right) \sin\left(\frac{n' \pi s'}{L'} \right) \\ & \times \left[ \frac{\delta_{ij} -\hat{k}_i \hat{k}_j}{c_T^5} e^{i |\x_0 - \x_0' +s \hat{n}(\sphericalangle) - s' \hat{n}'(\sphericalangle') | \omega \cos \theta /c_T } \right. \\ & \left.  \,\,\,\,\,\,\,\,\,\, + \frac{\hat{k}_i \hat{k}_j}{c_L^5} e^{i|\x_0 - \x_0' +s \hat{n}(\sphericalangle) - s' \hat{n}'(\sphericalangle') | \omega \cos \theta/c_L } \right]
\end{split}
\end{equation}
as a function of all the aforementioned variables. Here the unit vectors $\hat{n}, \hat{n}'$ point along the direction in which the string is laid out. The differential $d\Omega_+$ indicates to integrate over the upper hemisphere $\cos \theta > 0$ only, and $\hat{k}$ is the unit vector that spans the surface of the corresponding sphere. The integrals are less manageable than before because there is no particular alignment between the direction defined by $\x_0 - \x_0' +s \hat{n}(\sphericalangle) - s' \hat{n}'(\sphericalangle')$ and the local orientation matrices ${\bf M}_{ik}(\sphericalangle)$.

The complexity in dealing with this object resides in that the product of operators that we used earlier was a standard matrix product. However, now we have to deal with degrees of freedom that are continuous, and thus these products are, in general, operations of the sort
\beq
F_3(\x,\y) = \int_\z F_1(\x,\z) F_2(\z,\y),
\eeq
and thus the inverse operation appearing in the final scattering amplitude, $(1 - {\bf T}(\omega))^{-1}$, must be taken with respect to all the operations that are involved in taking products between effective propagators $F$. Finding analytic expressions for these objects is highly non-trivial in general.

Nonetheless, if we consider a situation with a finite number of dislocations, each with its own properties, the situation becomes somewhat more tractable: the products between ${\bf T}(\omega)$ again reduce to matricial products, in a vectorial space of dimension $n_D \times N$ with $N$ the number of dislocations in the solid. 
 
All computations is this section have been performed with the interaction~\eqref{quad-hamiltonian}, which is quadratic in the fields. We now go to the next level.

\section{Cubic interactions} \label{sec:cubic}

Even though the quadratic term in $S_{\rm int}$ fully captures the usual Peach-Koehler force, the fact that this comes from an action principle naturally suggests higher-order couplings. As a starting point, in this section we consider the cubic interactions, which give rise to scattering of phonons by strings with an energy transfer from one to the other. These are given by:
\beq
H_{I}^{(3)}(t) = \mu b \int_0^L \!\!\! ds \, {\bf M}_{kl} \frac{\partial^2 u_k^I}{\partial x_1 \partial x_l}((0,0,s),t) \left(X^I(s,t)\right)^2.
\eeq
Even though one could also study the corrections to the mass operator that arise from these terms, we shall postpone that discussion for section~\ref{sec:Quant}, where we will treat them inside a broader context of quantum-mechanical corrections.

For simplicity, we shall take $\x_0 = 0$ throughout this section, and denote $\omega_\iota \equiv \omega_{\iota}(\k)$ where a prime on the polarization index will also imply a prime on the wavenumber. Since the final purpose of this section will be to study in what regime one would expect the interactions of phonons by dislocations to compete with the cubic self-interactions of phonons, we will include the quadratic theory only to the lowest comparable order in the perturbative expansion that generates the processes of interest. The complete result, however, will include (at least in once per process) a $(1 - {\bf T}(\omega))^{-1}$ factor for each external particle (be it a phonon or a dislon) involved in the corresponding scattering process, accounting for the full extent of the quadratic interactions.

We now proceed to examine the processes of phonon scattering by excited strings and compare them to the effects of cubic elastic non-linearities. In between, we will take the opportunity to explore how the scattering cross-sections look like in the presence of a thermal distribution for one of the particles involved in each process.

\subsection{Scattering of a phonon by an excited string}

Consider the process of an ingoing elastic wave with wavenumber $\k'$ scattering with the string and producing an outgoing elastic wave with wavenumber $\k$, but this time accounting for an excitation of the string in the initial state. This is described by
\begin{equation}
\braket{f|i} = \braket{0 | a_{\iota}(\k) \text{T} \exp \left( - \frac{i}{\hbar} \int_{-\infty}^{\infty} \!\!\! dt \, H_I(t) \right) a_{\iota'}^{\dagger}(\k') a_n^{\dagger} | 0}.
\end{equation}
We can now repeat the process of computing the scattering amplitude as before by expanding the $S$-matrix in powers of $b$. The result is found to be given by $\braket{f|i} = 2\pi \, \delta(\omega_{\iota} - \omega_{\iota'} - \omega_n) \, \mathcal{T} $, with
\begin{equation} \label{disloc-2ph}
\begin{split}
\mathcal{T} = & -i E^*(\k;\iota) E(\k';\iota') \sqrt{\frac{\hbar}{(mL)^3 \omega_n}} \\ & \times \sum_{n'=1}^{n_D} \frac{4\pi^3 n'^3}{L^3}  \left[   \frac{(\k \cdot \hat{e}_1)/(\k \cdot \hat{e}_3)}{\omega_{n'}^2 - \omega_{\iota'}^2 }   \frac{e^{iL(\k' \cdot \hat{e}_3) - i n' \pi} - 1}{(\k' \cdot \hat{e}_3)^2 - \left(\frac{\pi n'}{L}\right)^2} \right. \\ & \,\,\,\,\,\,\,\,\,\,\,\,\,\,\,\,\,\,\,\,\,\,\,\,\,\,\,\,\,\,\,\,\,\,\, \times \left. \frac{e^{-iL(\k \cdot \hat{e}_3)} - 1}{(\k \cdot \hat{e}_3)^2 - \left(\frac{2\pi n'}{L}\right)^2}  + (\k \leftrightarrow - \k' ) \right].
\end{split}
\end{equation}
Just as the mass operator of the phonon exhibits resonances at the natural frequencies of the string, resonant scattering of a phonon by a string through cubic interactions is also possible. One only needs to notice the $(\omega_{n'}^2 - \omega_{\iota'}^2)$ factor in the denominator of the previous amplitude, which provides a distinctive signal to compare with other types of scattering. However, it is possible to argue that the exact result, in analogy with what we obtained for the quadratic theory, will have resonance peaks of finite width. 

Let us note that the opposite process, i.e. having a phonon lose energy to a string, is also possible and has a scattering amplitude with similar behavior. The corresponding amplitude is given by
\begin{equation}
\braket{f|i} = \braket{0 | a_n a_{\iota}(\k) \text{T} \exp \left( - \frac{i}{\hbar} \int_{-\infty}^{\infty} \!\!\! dt \, H_I(t) \right) a_{\iota'}^{\dagger}(\k')  | 0},
\end{equation}
and analogously the $\T$ matrix is given by
\begin{equation} \label{2ph-disloc}
\begin{split}
\mathcal{T} = & \, i E^*(\k;\iota) E(\k';\iota') \sqrt{\frac{\hbar}{(mL)^3 \omega_n}} \\ & \times \sum_{n'=1}^{n_D} \frac{4\pi^3 n'^3}{L^3}  \left[   \frac{(\k' \cdot \hat{e}_1)/(\k' \cdot \hat{e}_3)}{\omega_{n'}^2 - \omega_{\iota}^2 }   \frac{e^{-iL(\k \cdot \hat{e}_3) - i n' \pi} - 1}{(\k \cdot \hat{e}_3)^2 - \left(\frac{\pi n'}{L}\right)^2} \right. \\ & \,\,\,\,\,\,\,\,\,\,\,\,\,\,\,\,\,\,\,\,\,\,\,\,\,\,\,\,\,\,\,\,\,\,\, \times \left. \frac{e^{iL(\k' \cdot \hat{e}_3)} - 1}{(\k' \cdot \hat{e}_3)^2 - \left(\frac{2\pi n'}{L}\right)^2}  + (\k \leftrightarrow - \k' ) \right].
\end{split}
\end{equation}

Nonetheless, for the preceding computations to be of interest in a typical setting, we should compare the typical magnitude of the cross-section for phonon scattering by a string versus phonon scattering by phonons, given by the cubic term in the energy density of a continuous elastic medium.

\subsection{Phonon-phonon-phonon scattering}

Let us recall \cite{Landau1970} that the cubic term in the energy density of a continuous solid may be written as
\beq
H^{(3)}_{\rm elastic} = \int d^3x \, c_{pqmrst} \frac{\partial u_m}{\partial x_q}\frac{\partial u_p}{\partial x_r}\frac{\partial u_s}{\partial x_t}.
\eeq
Using the standard techniques employed so far, a quick computation for the amplitude of two incident phonons with wavenumber $\k_1, \k_2$ and an outgoing phonon with wavenumber $\k$ yields
\begin{equation}
\begin{split}
\braket{f | i}_{{\rm ph}^3} = & \, (2\pi)^4 \delta^{(4)}(k - k_1 - k_2) \sqrt{\frac{\hbar}{8\rho^3 \omega_{\iota} \omega_{\iota_1} \omega_{\iota_2}}} \\ & \,\,\,\,\,\,\,\, \times 6 c_{pqmrst} (\k \cdot \hat{e}_q) (\k_1 \cdot \hat{e}_r) (\k_2 \cdot \hat{e}_t) \\ & \,\,\,\,\,\,\,\, \times \varepsilon_{\iota}(\k)_m \varepsilon_{\iota_1}(\k_1)_p \varepsilon_{\iota_2}(\k_2)_s,
\end{split}
\end{equation}
where $ \delta^{(4)}(k - k_1 - k_2) = \delta(\omega_{\iota} - \omega_{\iota_1} - \omega_{\iota_2}) \delta^{(3)}(\k - \k_1 - \k_2)$ enforces overall energy-momentum conservation. In what follows, we will use $\mathcal{T}_{{\rm ph}^3}$ defined implicitly through $\braket{f | i}_{{\rm ph}^3} = (2\pi)^4  \delta^{(4)}(k - k_1 - k_2) \mathcal{T}_{{\rm ph}^3}$.

However, in order to make a comparison with the scattering amplitude by a string it is necessary to include a distribution of phonons to scatter against, as the dimensionality of the Dirac delta prevents us from making a direct comparison between the constants of the problem. Even though we could simply substitute it with a volume factor, this presents us with the opportunity to write down cross-sections through thermal distributions.

\subsection{Behavior at finite temperature}

Although our present interests encourage us to explore scattering of phonons through some distribution, the prospect of writing down the cross section of an incident wave on a solid at finite temperature $T$ should be rather attractive. Let us take an initial state characterized by a density matrix of thermodynamic equilibrium at chemical potential $\bar \mu$ in the grand canonical ensemble for the strings
\beq
\hat{\rho}^0 = \frac{a_{\iota'}^{\dagger}(\k') \! \ket{0} \! \bra{0} \! a_{\iota'}(\k')}{\braket{0|a_{\iota'}(\k') a_{\iota'}^{\dagger}(\k')|0} } \otimes \frac{e^{- \beta \sum_{n=1}^{n_D} (\hbar \omega_n - \bar \mu) a^{\dagger}_n a_n }}{{\rm tr} \left[e^{- \beta \sum_{n=1}^{n_D} (\hbar \omega_n - \bar \mu) a^{\dagger}_n a_n } \right] },
\eeq
where $\beta = (k_B T)^{-1}$. To make progress, it is helpful to write the exponential in terms of the string eigenstates:
\beq
\begin{split}
& e^{- \beta \sum_{n=1}^{n_D} (\hbar \omega_n - \bar \mu) a^{\dagger}_n a_n } \\ & =  \sum_{\{N_i\}_{i=1}^{n_D} } e^{- \beta \sum_{n=1}^{n_D} (\hbar \omega_n - \bar \mu) N_n } \! \ket{N_1, ...,N_{n_D}} \! \bra{N_1, ...,N_{n_D}}
\end{split}
\eeq
where the sum over $\{N_i\}_{i=1}^{n_D}$ represents the sum over all occupation numbers $N_i$ for each mode of the string $i \in \{1, ..., n_D \}$. Proceeding as we did before, the cross-section for outgoing phonons scattered through a process with scattering amplitude~\eqref{disloc-2ph} will be given by
\beq
d\sigma = \frac{d^3k}{c_{\iota'} (2\pi)^3}  \, {\rm tr} (U \hat{\rho}^0 U^{\dagger}  \! \ket{\k,\iota} \! \bra{\k,\iota} \otimes \mathcal{I}_{n_D} )
\eeq
which in turn implies
\beq
\begin{split}
d\sigma = \frac{d^3k}{c_{\iota'} (2\pi)^3} {\rm tr}_{\rm string} & \left[  \sum_{\{N_i\}_{i=1}^{n_D} } e^{- \beta \sum_{n=1}^{n_D} (\hbar \omega_n - \bar \mu) N_n } \right. \\ & \left. \times \bra{\k,\iota}  \! U \! \ket{N_1, ...,N_{n_D}} \otimes \ket{\k',\iota'} \right. \\ & \left. \times \bra{N_1, ...,N_{n_D}} \otimes \bra{\k',\iota'} \!  U^{\dagger}  \! \ket{\k,\iota} \right]
\end{split}
\eeq
and thus, if we only keep track of the processes that originate by~\eqref{disloc-2ph}, we get
\beq
\begin{split}
\frac{d\sigma}{d\Omega} = \sum_{\{N_i\}_{i=1}^{n_D} } & e^{- \beta \sum_{n=1}^{n_D} (\hbar \omega_n - \bar \mu) N_n } \\ & \times \sum_{i=1}^{n_D} N_i \frac{(\omega_{\iota'}+\omega_i)^2}{4\pi^2 c_{\iota'} c_\iota^3} |\mathcal{T}|^2_{\omega_\iota = \omega_{\iota'} + \omega_n}
\end{split}
\eeq
where the $N_i$ factor comes from selecting one of the creation operators in the states
\beq
\ket{N_1, ...,N_{n_D}} \equiv \frac{(a_{1}^{\dagger})^{N_1} \cdots (a_{n_D}^{\dagger})^{N_{n_D}} }{\sqrt{N_1! \cdots N_{n_D}!}} \ket{0}
\eeq
and normalizing appropriately when taking the trace.

Now one may execute the sum over the occupation numbers as usual in statistical mechanics. The resulting cross-section would be
\begin{equation} \label{cross-d2ph}
\frac{d\sigma}{d\Omega} = \sum_{n=1}^{n_D} \frac{|\mathcal{T}|^2_{\omega_\iota = \omega_{\iota'} + \omega_n} }{e^{(\hbar \omega_n - \bar \mu)/k_B T} - 1} \frac{(\omega_{\iota'} + \omega_n)^2}{4\pi^2 c_{\iota'} c_{\iota}^3}.
\end{equation}
with $\mathcal{T}$ given by \eqref{disloc-2ph}. Here we have assumed that the dislocations are in thermodynamic equilibrium with some reservoir at temperature $T$ and chemical potential $\bar \mu$, but not with the phonons. That is, we have neglected the effects of the interaction for this estimate. 

On the other hand, the scattering of phonons under similar assumptions gives
\begin{equation} \label{3ph}
\left. \frac{d\sigma}{d\Omega} \right|_{{\rm ph}^3}  =  \frac{|\mathcal{T}_{{\rm ph}^3}|^2_{\substack{ \k_2 = \k -\k_1 \\ \omega_{\iota_2} = \omega_\iota -\omega_{\iota_1}}}}{e^{\hbar (\omega_\iota - \omega_{\iota_1} )/k_B T} - 1} \frac{V \omega_\iota^2}{4\pi^2 c_{\iota_1} c_{\iota}^3} K(\k, \k_1;\{ c_{\iota_i}\}) 
\end{equation}
where $K(\k, \k_1; c_{\iota_i})$ is a kinematical factor arising from the restrictions imposed by energy-momentum conservation (details are given in Appendix~\ref{sec:kinematical}), and $V$ is the volume of the solid. Given that the volume of the solid is an extensive quantity that does not depend on the specific constitution of the solid, if there is only one dislocation present we could always consider a ``large volume'' limit and neglect the effects of the dislocation. Therefore, a meaningful comparison will arise only if we let the dislocations increase in number with the volume of the solid. Thus, the appropriate comparison between $\mathcal T$ matrices is $\T_{{\rm ph}^3}$ v/s $ \sqrt{n_d} \T$, where $n_d$ is the dislocation number density. 
 
To compare them, we will assume $c_{pqmrst}$ to be of the same order as $\mu$, and take a situation in which the incident and outgoing frequencies are not too close to the resonances $\omega_n$, so that \eqref{disloc-2ph} holds as written. This is because the exact result will contain higher order terms that after resummation should render the peaks at $\omega_n$ finite, in analogy with what happens in \eqref{exact-quadr}.

Given that the temperature enters the differential cross-sections in the same manner in both \eqref{cross-d2ph} and \eqref{3ph}, we can take the corresponding factors to be equal as long as we are interested in asymptotic states with the same ingoing and outgoing frequencies in both situations. A rough estimate gives
\beq
\mathcal{T}_{{\rm ph}^3} \sim \mu k^3 \sqrt{\frac{\hbar}{\rho^3 \omega_{\iota} \omega_{\iota_1} \omega_{\iota_2}}}
\eeq
and, assuming that the preponderant eigenfrequency of the string in the scattering process is $\omega_1$,
\beq
\mathcal{T} \sim \sqrt{\frac{\hbar n_d}{(mL)^3 \omega_{\iota} \omega_{\iota'} \omega_n}} \frac{\mu^2 b^2 L^2}{\rho} \frac{k^3}{\omega_1^2}
\eeq
one obtains that the scattering by excited strings should dominate provided that 
\beq
n_d L^3 \gg b^2/L^2.
\eeq

Therefore, if there is a high density of scatterers compared to the ``aspect ratio'' of the dislocations, it would be expectable that string dislocations represent the dominant contribution to the phonon cross-section. For example, if we take a Burgers vector $b=0.5$ nm and a dislocation length $L=50$ nm, we have that phonon-dislocation scattering will dominate over phonon-phonon scattering for dislocation densities $\Lambda = n_d L \sim 10^8$ cm$^{-2}$, which is, in the opinion of the authors, a very modest bound. 

All of the previous results could have been obtained through a careful classical analysis of the original action integral $S$. Now we turn to some intrinsically quantum mechanical effects that may be reflected on different observable quantities.

\section{Quantum corrections} \label{sec:Quant}

As was mentioned earlier, the action integral $S_{\rm int}$ discussed prior to section~\ref{sec:cubic} only includes quadratic terms. However, the starting point is
\begin{equation}
S_{\rm int} = - b_i \int dt \int_{\mathcal{S}} dS^j \sigma_{ij}.
\end{equation}
If we perform a Taylor expansion of $\sigma_{ij}(s\hat{\tau} + r \vec{X})$ about the equilibrium position of the string $r=0$ ($r, s$ being the coordinates parametrizing $\delta \mathcal{S}$), we obtain
\begin{equation} \label{full-S-int}
\begin{split}
S_{\rm int} &= - b_i \int dt \int_{\mathcal{S}} dS^j \left[ \sum_{n=0}^{\infty} c_{ijkl} \frac{\partial^{n+1} u_k}{\partial x_1^n  \partial x_l} \frac{r^n}{n!} X^n \right] \\ &= -  \mu b {\bf M}_{kl} \int dt \int_0^L \!\! ds  \left[ \sum_{n=0}^{\infty} \frac{\partial^{n+1} u_k}{\partial x_1^n  \partial x_l} \frac{X^{n+1}}{(n+1)!} \right],
\end{split}
\end{equation}
which gives further couplings between phonons and string dislocations. These terms will make a phonon be able to interact with several modes of the dislocation in a single scattering event at the same time.

For the moment we will only deal with the effects of this additional terms in action integral on the scattering of phonons by a string to lowest order in the amplitude of the interaction, which schematically would be $\mathcal{O}(b^2)$ (as we have shown earlier that this is not the true perturbative parameter, at least in the quadratic theory). The main difference of this section with our earlier computations is that now the internal structure of the scattering will get richer. Namely, ``loop'' contributions, as one would call them in quantum field theory, will emerge when performing ``contractions'' (i.e. pairings through Wick's theorem when computing the quantum expectation value) of fields in the interaction hamiltonians with themselves.

\subsection{The phonon-phonon amplitude}

The object of interest is the same as in section~\ref{sec:quad}, and given by equation~\eqref{ph-ph-a}
\beq 
\langle f|i\rangle = \braket{0| a_{\iota}(\k) \text{T} \exp \left[ - \frac{i}{\hbar} \int_{-\infty}^{\infty} H_I(t) dt \right] a_{\iota'}^{\dagger}(\k')  |0}, \nonumber
\eeq
only that now we have a more complex hamiltonian $H_I$. However, it is still linear in $\u$, and therefore the factors $E(\k;\iota)$ are still present. This only leaves the aforementioned ``loop'' contributions yet to be accounted for. If $n,n'$ denote indices in the sum of the interaction picture hamiltonian derived from~\eqref{full-S-int}, we are left with terms proportional to
\beq \label{pairings}
\frac{(-i\k \cdot \hat{e}_1)^n}{(n+1)!} \frac{(+i\k' \cdot \hat{e}_1)^{n'}}{ (n'+1)!} \braket{ 0 | {\text T} X_I^{n+1}(s,t) X_I^{n'+1}(s',t') | 0 },
\eeq
in which the unraveling of the products of creation and annihilation operators may be performed as a combinatorial exercise. After doing so, the expressions may be summed back to obtain
\begin{equation} \label{quantum-b2}
\begin{split}
\mathcal{T} =&  - \frac{E^{*}(\k;\iota) E(\k'; \iota')}{ \hbar (\k \cdot \hat{e}_1) (\k' \cdot \hat{e}_1)} \int_0^L \!\!\! ds \int_0^L \!\!\! ds' \!  \int_{-\infty}^{\infty} \!\!\!\! dt  \\ & \,\,\,\,\,\,\,\,\,\,\,\,\,\,\,\,\,\,\,\,\, \times e^{   - i(\k \cdot \hat{e}_3) s  - (\k \cdot \hat{e}_1)^2 \Delta(s)/2 }    \\ & \,\,\,\,\,\,\,\,\,\,\,\,\,\,\,\,\,\,\,\,\,  \times  e^{ +  i\omega t  } \left[e^{ (\k \cdot \hat{e}_1) (\k' \cdot \hat{e}_1) \Delta(s,s',t) } - 1 \right] \\
  & \,\,\,\,\,\,\,\,\,\,\,\,\,\,\,\,\,\,\,\,\, \times e^{    i(\k' \cdot \hat{e}_3) s'  - (\k' \cdot \hat{e}_1)^2 \Delta(s')/2 }
\end{split}
\end{equation}
where we have assumed $\omega \neq 0$, defined the $\mathcal T$ matrix through $\braket{f|i}|_{\mathcal{O}(b^2)} =  2\pi \delta(\omega_{\iota}(\k) -\omega_{\iota'}(\k')) \mathcal{T} $, and further denoted
\beq
\Delta(s) \equiv \Delta(s,s,0).
\eeq
We point out to the interested reader the existence of general formulae in quantum field theory~\cite{ScheihingHitschfeld:2019abs} that allow to infer this result by looking at the analytic structure of the interaction hamiltonian. The reason why this occurs is that the stress tensor $\sigma_{ij}$ appearing in the Lagrangian density may be formally written as
\bea
\sigma_{ij}(s\hat{\tau} + r \vec{X}) &=& \exp \left[ r \vec{X} \cdot \nabla_{\x} \right] \sigma_{ij}(s\hat{\tau} + \x)|_{\x = 0} \nonumber \\
&=& \left[ \exp \left( r \vec{X} \cdot \nabla_{\x} \right) c_{ijkl} \frac{\partial u_k}{\partial x_l}(\x) \right]_{\x = s\hat{\tau}}
\eea
and then in momentum space the gradient takes the form of a wavenumber, thus showing that the exponentials contained in \eqref{quantum-b2} are accounting for precisely this structure. These exponentials are preserved along the result because performing the contractions of equation~\eqref{pairings} is equivalent to computing an expectation value over a gaussian probability distribution, and thus the outcome naturally involves exponentials of the propagators.

It is important to stress that equation~\eqref{quantum-b2} contains the full extent of the quantum corrections to the phonon-phonon scattering amplitude implied by the action integral~\eqref{full-S-int}, to the lowest order in the amplitude of the interaction, as the propagators in the exponential are proportional to $\hbar$. Indeed, one may treat~\eqref{quantum-b2} as a power series on $\hbar$, where the zeroth-order term gives the contribution of the quadratic theory and the subsequent terms reveal quantum mechanical effects. It is interesting to note that at this point a new characteristic length emerges: the product of quantities that give the string propagator its dimensions. Simple inspection gives that
\beq
d_q \equiv \sqrt{\frac{\hbar}{\rho b^2 c_T}} \sim \sqrt{\frac{\hbar}{m L \omega_1}}
\eeq
plays a similar role to $L$ regarding how large the corrections are with respect to the first order computation. If we plug in typical numerical values for $\rho$, $b$, and $c_T$, for instance, those of silicon, then we find $d_q \sim 10^{-9}$ cm.

In the spirit of alleviating the discussion while trying to keep track of the quantum effects that emerge in the presence of nonlinear interactions with dislocations, we will now study the quantum corrections explicitly in the case $n_D = 1$, and to first order in $\hbar$ for the general case.

\subsection{The $n_D = 1$ case: an analytic exploration}

Following the approach we have taken thus far, studying a $n_D = 1$ dislocation segment is equivalent to studying a string with a single oscillation mode, which may represent a very short dislocation in units of the interatomic distance of the crystal. Conversely, imposing it should always give a reasonable description of the physics at frequencies below $c_T/L$, but since we are currently exploring quantum corrections, we find the former to be the most conservative attitude towards this setting.

Nonetheless, it is the tractability of the computations at hand that turns our attention towards this particular case. For instance, the string propagators take simpler forms:
\bea
\Delta(s) &=& \frac{\hbar}{mL \omega_1} \sin^2 \left( \frac{\pi s}{L} \right) \\
\Delta(s,s',t) &=& \frac{\hbar}{mL } \frac{e^{-i\omega_1|t|}}{\omega_1} \sin \left( \frac{\pi s}{L} \right) \sin \left( \frac{\pi s'}{L} \right),
\eea
and out of convenience we shall define
\beq
\Delta(s,s') \equiv \frac{\hbar}{mL \omega_1} \sin \left( \frac{\pi s}{L} \right) \sin \left( \frac{\pi s'}{L} \right).
\eeq
Using these expressions, the temporal integral in~\eqref{quantum-b2} may be carried out explicitly in terms of incomplete Gamma functions. The relevant result to this is
\beq \label{fourier-integrated}
\begin{split}
& \int_{-\infty}^{\infty} dt \, e^{i\omega t} e^{k_1 k_1' \Delta(s,s',t)} \\ &= \frac{- i}{4\pi \omega} \left[  (- k_1 k_1' \Delta(s,s') )^{-\frac{\omega^2}{\omega_1^2}} \gamma \! \left(1 + \frac{\omega^2}{\omega_1^2}, -k_1 k_1' \Delta(s,s')\right) \right. \\ & \,\,\,\,\,\,\,\,\,\,\,\,\,\,\,\,\,\, \left.  - (-k_1 k_1' \Delta(s,s') )^{\frac{\omega^2}{\omega_1^2}} \gamma \! \left(1 - \frac{\omega^2}{\omega_1^2}, -k_1 k_1'\Delta(s,s')\right) \right],
\end{split}
\eeq
where $\gamma(a,x)$ is the lower incomplete Gamma function, which admits an analytic expansion
\beq
\gamma(a,x) = x^a \Gamma(a) e^{-x} \sum_{n=0}^{\infty} \frac{x^n}{\Gamma(a+n+1)}.
\eeq
Then, what remains is an integral over a bounded domain of an analytic function, which may be solved numerically for the values of interest without issue. However, we will not proceed to do so since for values of practical interest ($d_q \sim 10^{-11}$ m) we can carry out a perturbative computation without losing much accuracy.

Before concluding this discussion, there is one feature of interest that is worth pointing out: by examining~\eqref{fourier-integrated}, we see that the result has poles in the frequency variable $\omega$ for each positive multiple of $\omega_1$ (i.e., $\omega = n\omega_1$, with $n$ a positive integer). Therefore, we expect that the scattering cross-section will have resonances at every frequency that is a multiple of $\omega_1$, although their amplitude will be suppressed by factors of $(k d_q)^2$ for every phononic excitation.

In the following subsection we give a concrete estimate of this corrections, for the general $n_D$ case.

\subsection{General $\mathcal{O}(\hbar)$ corrections}

For practical purposes, since most of the time we will have $kd_q \ll 1$, in order to estimate the corrections to the scattering amplitude due to quantum effects it is sufficient to keep the first correction to the result of the quadratic theory, which is proportional to $\hbar$. As the difficulty of the computation is significantly reduced with respect to that of the exact quantity, we can afford to leave $n_D$ as arbitrary.

Performing a Taylor expansion on $\hbar$, we can approximate~\eqref{quantum-b2} with
\bea \label{quantum-first-order}
\mathcal{T} &=&  - \frac{E^{*}(\k;\iota) E(\k'; \iota')}{ \hbar k_1 k_1'} \int_0^L \!\!\! ds \int_0^L \!\!\! ds' \!  \int_{-\infty}^{\infty} \!\!\!\! dt \nonumber  \\ & & \times e^{   - i k_3 s } \left[ 1 - \frac{k_1^2}{2} \Delta(s) \right]  e^{    ik_3' s'} \left[ 1  -  \frac{k_1'^2}{2} \Delta(s') \right] \nonumber \\   & &  \times  e^{ +  i\omega t  } \left[ k_1 k_1' \Delta(s,s',t) + \frac{k_1^2 k_1'^2}{2} \Delta(s,s',t)^2 \right] \nonumber \\
&=& \frac{2 i L E^{*}(\k;\iota) E(\k'; \iota')}{ m} \left[ \sum_{n=1}^{n_D} \frac{f^{(0)}_n(-k_3) f^{(0)}_n(k_3') }{\omega_n^2 -\omega^2 } \right. \nonumber \\ & & \left. + \frac{\hbar k_1k_1'}{2 mL \omega_1} \sum_{n,n'=1}^{n_D} \frac{n+n'}{nn'} \frac{f^{(1)}_{n,n'}(-k_3) f^{(1)}_{n,n'}(k_3') }{(\omega_n + \omega_{n'})^2 -\omega^2 } \right. \nonumber \\ & & \left. - \frac{\hbar k_1^2}{2mL \omega_1} \sum_{n,n'=1}^{n_D} \frac{1}{n'} \frac{f^{(2)}_{n,n'}(-k_3) f^{(0)}_{n}(k_3') }{\omega_n^2 -\omega^2 }  \right. \nonumber \\ & & \left. - \frac{\hbar k_1'^2}{2mL \omega_1} \sum_{n,n'=1}^{n_D} \frac{1}{n'} \frac{f^{(0)}_{n}(-k_3) f^{(2)}_{n,n'}(k_3') }{\omega_n^2 -\omega^2 } \right],
\eea
where the functions $f^{(i)}_{j(,k)}(k_3)$ are defined in Appendix~\ref{sec:quant-form}. This constitutes the lowest order correction in $\hbar$ to the scattering amplitude defined by the quadratic theory (at its respective lowest order in perturbation theory).

For our present purposes, their most important property is that they peak when $k_3 L$  is a specific integer multiple of $\pi$. Furthermore, at those points they take values of order $\mathcal{O}(1)$. For instance, $f_n^{(0)}$ peaks in amplitude at $k_3 L \sim n\pi$, while $f_{n,n'}^{(1)}$ has peaks at $k_3 L \sim (n + n')\pi$. 

This makes it possible to estimate, approximately, the precise order of magnitude of the quantum correction to the scattering amplitude: we need only compare the factors accompanying each term of the form $(\omega_n^2 - \omega^2)$, without worrying about divergences at $\omega = \omega_n$ as the exact computation will sum them into a finite expression. By rearranging the sum inside the square brackets in~\eqref{quantum-first-order} with prefactor $\frac{\hbar k_1 k_1'}{2mL \omega_1}$ into a sum over $n + n'$ and $n'$, we may take $f_{n,n'}^{(1)}$ as weakly dependent on $n'$ because its peak value close to $kL \sim (n + n') \pi$ will remain nearly constant (i.e., roughly independent of $n'$ as long as $n+n'$ remains fixed), preserving the dominant behavior as a function of $k$. Thus, we are left with harmonic sums that we can approximate by the means of $\sum_{n'=1}^{n^*} \frac{1}{n'} \approx \gamma_e + \ln (n^*)$, with $\gamma_e$ the Euler-Mascheroni constant.

Therefore, we find that the peak of the scattering amplitude at the resonance frequency $\omega_{n^*}$ should be subject to corrections of magnitude
\beq
\sim \frac{\hbar k_1 k_1'}{2mL \omega_1} \left(\gamma_e + \ln(n^*-1) \right)
\eeq
as a consequence of the term with $f_{n,n'}^{(1)}$ factors, while the ones with $f_{n,n'}^{(2)}$ induce a correction of size
\beq
\sim \frac{\hbar (k_1^2 + k_1'^2) }{4mL \omega_1} \left( \gamma_e + \ln(n_D) \right)
\eeq
to each resonance peak. Consequently, as the dependence on $n^*$ and $n_D$ is logarithmic, a short wavelength phonon with $k \sim 2\pi \cdot 10^9 \, \text{m}^{-1}$ will have its scattering amplitude corrected by a factor of roughly
\beq
1 - \pi^2 \times 10^{-4} \times \left( \gamma_e + \ln(n_D) \right),
\eeq
where we have taken $\gamma = c_L/c_T \sim 2$ and $\ln (\delta/\delta_0) \sim 3 $.

In terms of measurability, we would like $n_D$ to be as large as possible. Recent experiments~\cite{Sun2018} show that it may be possible to probe samples with dislocation lengths up to $L \sim 1 \, \mu \text{m}$, which sets $\ln (n_D)$ to be at most $\sim 10$. 

Thus, we deduce that in an optimistic scenario for the detectability of quantum corrections, those effects will be of relative size $1 \%$ to the classical predictions. While this may seem far-fetched given that one would have to make a controlled experiment with phonons of wavelength comparable to the lattice constant in a carefully devised material with dislocations of mesoscopic length while assuming the continuum model for the string still holds, the result doesn't vary much with $n_D$: should $L$ be of order $\sim 10 \, \text{nm}$, the correction would only get reduced by a factor of $1/2$. Therefore, the greatest challenge to test these predictions is to measure the scattering amplitude for phonons with high wavenumber.

\section{Discussion and {outlook}}
\label{discussion}

In this paper we have studied a theory of phonons in an elastic continuum interacting with quantum dislocations segments. The results contained herein establish, to the best of the authors' knowledge, the first time that such a theory incorporating the Peach-Koehler force has been developed. In addition to that, for the quadratic action we established the finiteness of the theory at all wavelengths, avoiding the need to engage in renormalization procedures previously used in classical settings\cite{Churochkin2016}.

{
The findings laid out in this paper can be used and generalized in a variety of ways. First would be the replacement of a continuum solid by a crystal lattice. The free phonon aspects of the theory would be straightforward. On the other hand, the free dislocation term in the action would have to be considered with some care, taking into account the actual lattice that is considered, particularly its slip systems. Moreover, the interaction term would have to be worked out in a consistent fashion. Consideration of screw, rather than edge, dislocations, should be a straightforward matter.
}

{
The theory of this paper has been developed in three dimensions. It should not be difficult to specialize the action (\ref{actionFL}) to two dimensions, in which dislocations are points rather than strings so the whole formalism should be easier to manage. Indeed, this could be a good place to start working out the consequences of having a crystal lattice rather than a continuum solid. It could also have significant consequences for a number of materials of current interest \cite{Tan2018,Gu2018}.
}

Another direction of research is to study this model in the $L \to \infty$ limit, along with $n_D \to \infty$. While some qualitative features of this limit have been explored throughout this paper, it would be interesting to start with a translationally invariant action along the equilibrium axis of the dislocation. The features observed in this limit should be sharper, as momentum conservation would be enforced through the aforementioned symmetry. Moreover, the length scale $L$ would disappear from the results, leaving a scale invariant theory where in all likelihood the scattering amplitudes will only depend on geometric features of the problem (angle of incidence, polarization vectors).

{
We have considered in some detail the scattering of phonons by quantum dislocations. It should be possible to use these results, in conjunction with, say, kinetic theory arguments, to compute the contribution of quantum dislocations segments to thermal conductivity. Results such as Eqn. (\ref{exact-quadr}) for the ${\cal T}$ matrix indicate that high frequency phonons can have a significant contribution, contrary to the classical intuition that only phonons of wavelength comparable to the dislocation length $L$, hence low frequency, would contribute.
}

Finally, we took into account the intrinsically quantum-mechanical effects that would single out the interaction Hamiltonian among others that could generate the Peach-Koehler force at leading order. These corrections are small even for phonons of wavelengths comparable to the lattice constant, but at each order in perturbation theory include distinct form factors that might render the signal identifiable if detected. 

{
On a longer term perspective, it should be possible to study the interaction of quantum dislocation segments with electrons. Even at the level of a quadratic theory, the nontrivial interaction studied in this work should give rise to distinct correlations between electrons and phonons if the electron-dislon coupling is included.
}

\begin{acknowledgments}

BSH is supported by a CONICYT grant number CONICYT-PFCHA/Mag\'{i}sterNacional/2018-22181513. FL gratefully acknowledges the support of Fondecyt Grant 1160823. BSH wishes to thank Luis E. F. Fo\`{a} Torres for helpful discussions.

\end{acknowledgments}

\begin{appendix}

\section{Averages over internal degrees of freedom: the quadratic theory} \label{sec:angular}

If we have a solid filled with randomly oriented dislocations, then it is expectable that in the presence of weak scattering the amplitudes and cross-sections are well described by averaging over internal degrees of freedom. Namely, to start with a definite object, we would like to average equation~\eqref{T-lowest}, which we now write in full in terms of the basic quantities of the problem
\begin{widetext}
\begin{equation} \label{T-lowest-a}
\begin{split}
\mathcal{T} = & \, \frac{ i \mu^2 b^2 e^{i \x_0 \cdot (\k' - \k)} k_l {\bf M}_{kl} \varepsilon_\iota(\k)_k  k'_{l'} {\bf M}_{k'l'}  \varepsilon_{\iota'}(\k')_{k'} }{ m L \rho \sqrt{\omega_{\iota}(\k) \omega_{\iota'}(\k')} }   \\  &\times \sum_{n=1}^{n_D} \sum_{n'=1}^{n_D}  \frac{n\pi/L}{\sqrt{\omega_n^2 - \omega^2}}  \frac{e^{-iL(\k \cdot \hat{e}_3) + i n \pi} - 1}{(\k \! \cdot \! \hat{e}_3)^2 - (n\pi/L)^2} \left[\frac{1}{1 - {\bf T}(\omega)} \right]_{nn'}  \frac{e^{iL(\k' \cdot \hat{e}_3) - i n' \pi} - 1}{(\k' \! \cdot \! \hat{e}_3)^2 - (n'\pi/L)^2} \frac{n'\pi/L}{\sqrt{\omega_{n'}^2 - \omega^2}},
\end{split}
\end{equation}
\end{widetext}
where summation over repeated indices $k,k',l,l'$ is implied. The first, and easiest, task to carry out is to perform a spatial average on the distribution of dislocations. This translates into integrating the previous equation over $\x_0$ and multiplying by $n_d$ (the dislocation number density): 
\beq
\bar{\T}^{\rm spatial} = n_d \int_{\x_0} \T|_{\x_0} = n_d \delta^{(3)}(\k - \k') \times \T(\k=\k').
\eeq
Then, all that remains is to compute an angular average, i.e. over the possible orientations of the triad $\hat{e}_1, \hat{e}_2, \hat{e}_3$.

At this point we note that $\hat{k}$ is the only physically relevant spatial direction remaining after the average is performed. Thus, for the purposes of computing an angular average, we may define integration angles around this axis. Let $\hat{i}, \hat{j}, \hat{k}$ be the basis of reference, with $\hat{i}, \hat{j}$ fixed unit vectors orthogonal to $\hat{k}$. Explicitly, if we let $(\theta,\varphi)$ be the spherical coordinate angles with respect to $\{\hat{i}, \hat{j}, \hat{k} \}$ and denote by $\{\hat{r}, \hat{\theta}, \hat{\varphi}\}$ the corresponding unit vectors, 
\bea
\hat{e}_1 &=& \cos \phi \, \hat{\theta} + \sin \phi \, \hat{\varphi} \\
\hat{e}_2 &=& \cos \phi \, \hat{\varphi} - \sin \phi \, \hat{\theta} \\
\hat{e}_3 &=& \hat{r}.
\eea
These conventions allow to represent the direction in which the string is laid out through the angles $(\theta,\varphi)$ and the relative orientation of the Burgers vector through the angle $\phi$. Therefore, an angular average of a function $f$ of the angles $(\theta,\varphi,\phi)$ is given by
\beq
\bar{f} = \frac{1}{8\pi^2} \int_0^{2\pi} \!\!\!\! d\phi \int_0^{2\pi} \!\!\!\!  d\varphi \int_0^{\pi} \!\!  d\theta \sin \theta f(\theta, \varphi, \phi).
\eeq

Now we proceed to compute the averaged scattering amplitude $\T$. The object of interest will be
\begin{widetext}
\begin{equation} \label{average-1}
\frac{1}{8\pi^2} \int_0^{2\pi} \!\!\!\! d\phi \int_0^{2\pi} \!\!\!\!  d\varphi \int_0^{\pi} \!\!  d\theta \sin \theta   k_l  k_{l'} {\bf M}_{kl}   {\bf M}_{k'l'}  \frac{e^{-iL(\k \cdot \hat{e}_3) + i n \pi} - 1}{(\k \! \cdot \! \hat{e}_3)^2 - (n\pi/L)^2}  \frac{e^{iL(\k \cdot \hat{e}_3) - i n' \pi} - 1}{(\k \! \cdot \! \hat{e}_3)^2 - (n'\pi/L)^2},
\end{equation}
leaving the polarization vectors out of the discussion so as to obtain a rank-2 tensor as a result and thus be able to choose the polarization modes later. In terms of the angles we defined earlier, equation~\eqref{average-1} is written as
\beq
\begin{split}
\frac{k^2}{8\pi^2} \int_0^{2\pi} \!\!\!\! d\phi \int_0^{2\pi} \!\!\!\!  d\varphi \int_0^{\pi} \!\!  d\theta \sin \theta & \left( (\cos \phi \, \hat{\theta}_{k} + \sin \phi \, \hat{\varphi}_{k} )(\cos \phi \, \hat{\theta}_{k'} + \sin \phi \, \hat{\varphi}_{k'})  \sin^2\theta \sin^2\phi \right. \\ 
& \left. - (\cos \phi \, \hat{\theta}_{k} + \sin \phi \, \hat{\varphi}_{k})(\cos \phi \, \hat{\varphi}_{k'} - \sin \phi \, \hat{\theta}_{k'}) \sin^2\theta \sin\phi \cos\phi  \right. \\ 
& \left.  -  (\cos \phi \, \hat{\varphi}_{k} - \sin \phi \, \hat{\theta}_{k})(\cos \phi \, \hat{\theta}_{k'} + \sin \phi \, \hat{\varphi}_{k'}) \sin^2\theta \sin\phi \cos\phi \right. \\ 
 & \left. + (\cos \phi \, \hat{\varphi}_{k} - \sin \phi \, \hat{\theta}_{k})(\cos \phi \, \hat{\varphi}_{k'} - \sin \phi \, \hat{\theta}_{k'}) \sin^2 \theta \cos^2\phi \right) \\ 
 & \times \frac{e^{-ikL\cos \theta} (-1)^n - 1}{(k \cos \theta)^2 - (n\pi/L)^2}  \frac{e^{ikL\cos \theta}(-1)^{n'} - 1}{(k \cos \theta)^2 - (n'\pi/L)^2},  
\end{split}
\eeq
and now the trigonometric integrals over $(\varphi, \phi)$ are straightforward. Performing them gives
\beq
\frac{k^2}{8} \int_0^\pi d\theta \sin^3 \theta \left(2 \sin^2 \theta \, \hat{k}_k \hat{k}_{k'} + (1 + \cos^2 \theta) (\delta_{kk'} -\hat{k}_k \hat{k}_{k'} ) \right) \frac{e^{-ikL\cos \theta} (-1)^n - 1}{(k \cos \theta)^2 - (n\pi/L)^2}  \frac{e^{ikL\cos \theta}(-1)^{n'} - 1}{(k \cos \theta)^2 - (n'\pi/L)^2}, 
\eeq
and changing variables to $u = \cos \theta$ we get
\beq \label{angular-done}
\frac{k^2}{8} \int_{-1}^1 du  \left( 2 (1-u^2)^2 \hat{k}_k \hat{k}_{k'} + (1 - u^4) (\delta_{kk'} -\hat{k}_k \hat{k}_{k'} ) \right) \frac{e^{-ikLu} (-1)^n - 1}{(k u)^2 - (n\pi/L)^2}  \frac{e^{ikLu}(-1)^{n'} - 1}{(k u)^2 - (n'\pi/L)^2}, 
\eeq
\end{widetext}
which upon substitution in $\bar{\T}$ yields the coefficients ${\bf F}_T(k,\omega), {\bf F}_L(k,\omega)$. 

\section{Phonon scattering by dislocation to all orders in the quadratic coupling} \label{sec:details}

Let us note that given a quadratic interaction Hamiltonian, the computation of expectation values acquires a simple form: by expanding~\eqref{ph-ph-a} in a power series of $H_I$, one needs to evaluate
\beq \label{schem-inter}
\braket{0 | a_{\iota'}(\k') H_I(t_1) \cdots H_I(t_n)  a_{\iota}^{\dagger}(\k) | 0},
\eeq
which can be solved through repeated use of the commutation relations for $a, a^{\dagger}$ and $a\ket{0} = 0$. Since every $H_I$ is quadratic, and more precisely bilinear on $\u$ and $X$, each interaction hamiltonian can be commuted exactly one time with $a_{\iota}(\k)/a_{\iota}^{\dagger}(\k)$ and exactly one time with $a_n/a_n^{\dagger}$. This means that one can build sequences, which are often written in terms of \textit{Feynman diagrams} in the standard approach to QFT, to represent each contribution to the scattering amplitude. Even though closed sequences (loops) can appear, they arise in every scattering amplitude (even in the vacuum-vacuum amplitude) as a multiplicative factor, and thus they can be treated as overall normalization to the observables.

Therefore, we can restrict ourselves to studying the fully connected contributions only (i.e. those that cannot be separated into products of scattering amplitudes of simpler processes), and evaluate only the sequences in~\eqref{schem-inter} that start with $a_{\iota'}(\k')$ and end with $a_{\iota}^{\dagger}(\k)$. Furthermore, using time-ordered quantities reveals that every sequence is equivalent, and they may all be summed together to obtain a compact result. Namely, if we define
\begin{widetext}
\begin{equation}
\Delta_{kl}(\k,s,t) \equiv \braket{0 |a_{\lambda}(\k) \frac{\partial u_k^I}{\partial x_l}((x_0,y_0,z_0+s),t) |0} = \sqrt{\frac{\hbar}{\rho}} \varepsilon_{\lambda}(\k)_k (-ik_l)  e^{-ik_3s} \frac{e^{i\omega_{\lambda}(\k)t }}{\sqrt{2 \omega_{\lambda}(\k)}} e^{-i \x_0 \cdot \k},
\end{equation}
\begin{equation}
\Delta_{klk'l'}(s-s',t-t') \equiv \Delta_{klk'l'}((x_0,y_0,z_0+s)-(x_0,y_0,z_0+s'),t-t'),
\end{equation}
then we may write
\begin{equation}
\begin{split}
\braket{f|i}|_{\mathcal{O}(b^{2n})} =  \left(-\frac{b^2}{\hbar^2}\right)^n \int_{-\infty}^{\infty} \!\!\!\! dt_1 \int_0^L \!\!\! ds_1 \, ... & \int_{-\infty}^{\infty} \!\!\!\! dt_{2n} \int_0^L \!\!\! ds_{2n} \Delta_{k_1 l_1}(\k^f,s_1,t_1) c_{12k_1l_1} \Delta(s_1,s_2,t_1-t_2) c_{12k_2l_2}   \\  \times & \Delta_{k_2l_2k_3l_3}(s_2-s_3,t_2-t_3) c_{12k_3l_3} \cdots c_{12k_{2n-2}l_{2n-2}} \\  \times &  \Delta_{k_{2n-2}l_{2n-2}k_{2n-1}l_{2n-1}}(s_{2n-2}-s_{2n-1},t_{2n-2}-t_{2n-1})  c_{12k_{2n-1}l_{2n-1}}  \\  \times & \Delta(s_{2n-1},s_{2n},t_{2n-1}-t_{2n}) c_{12k_{2n}l_{2n}} \Delta_{k_{2n}l_{2n}}^*(\k^i,s_{2n},t_{2n}).
\end{split}
\end{equation}
\end{widetext}
Now we use the known formula
\begin{equation} \label{known-res}
\frac{e^{-i \omega |t-t'|}}{2\omega}  = \frac{1}{2 \pi i} \int_{-\infty}^{\infty} d\omega' \frac{e^{-i\omega'(t-t')}}{-\omega'^2 + \omega^2 - i0^{+}},
\end{equation}
where $0^{+}$ is the distributional positive infinitesimal that defines the prescription to be used when evaluating the integral, usually denoted by $\epsilon$. With this expression we can perform explicitly the time integrals and get an overall energy conservation factor. It turns out that the integrals over the string coordinate $s$ are also easy to do (this is because for each term in the string propagator, the $s$ and $s'$ contributions are separable within each sum): if we define
\begin{widetext}
\begin{equation} \label{mixed-prop-a}
\begin{split}
F(n,n') \equiv \frac{ \mu^2 b^2 {\bf M}_{kl} {\bf M}_{k'l'}}{ \rho} \int_0^L \!\!\! ds \int_0^L \!\!\!  ds' & \sin\left(\frac{n \pi s}{L} \right) \sin\left(\frac{n' \pi s'}{L} \right) \int \frac{d^3k}{(2\pi)^3} e^{i (\k \cdot \hat{e}_3) (s - s') }  \\ & \times k_l k_{l'}  \left[ \left(\delta_{kk'} - \frac{k_k k_{k'}}{k^2} \right)\frac{1}{\omega_T(\k)^2 - \omega^2 - i0^{+}} + \frac{k_k k_{k'}}{k^2} \frac{1}{ \omega_L(\k)^2 - \omega^2 - i0^{+}} \right],
\end{split}
\end{equation}
then we find that
\begin{equation}
\begin{split}
\mathcal{T}|_{\mathcal{O}(b^{2N})} = i  E^*(\k;\iota)  \sum_{n_{1}=1}^{n_D} \frac{e^{-iL(\k \cdot \hat{e}_3) + i n_{1} \pi} - 1}{(\k \cdot \hat{e}_3)^2 - (n_{1}\pi/L)^2} \frac{2 n_1 \pi/L }{mL (\omega_{n_{1}}^2 - \omega^2 )}  \sum_{n_{2}=1}^{n_D} F(n_{1}, n_{2}) \frac{2 }{mL (\omega_{n_{2}}^2 - \omega^2 )} \\
\sum_{n_{3}=1}^{n_D} F(n_{2}, n_{3}) \frac{2 }{mL (\omega_{n_{3}}^2 - \omega^2 )} \cdots \sum_{n_{N}=1}^{n_D} F(n_{N-1}, n_{N}) \frac{2 n_N \pi/L}{mL (\omega_{n_{N}}^2 - \omega^2 )}  \frac{e^{iL(\k' \cdot \hat{e}_3) - i n_N \pi} - 1}{(\k' \cdot \hat{e}_3)^2 - (n_N\pi/L)^2} E(\k';\iota') .
\end{split}
\end{equation}
\end{widetext}
This can be written in a more succinct manner: if we define the following linear operator ${\bf T}(\omega)$ through its matrix elements
\beq
[{\bf T}]_{n,n'}(\omega) \equiv \frac{2}{mL} \frac{F(n, n')}{\sqrt{ \omega_{n}^2 - \omega^2 } \sqrt{ \omega_{n'}^2 - \omega^2 }} 
\eeq
and the $n_D$-dimensional vectors
\beq
[v(\k)]_n \equiv \sqrt{\frac{2 (n \pi/L )^2}{mL (\omega_{n}^2 - \omega^2 )}}  \frac{e^{iL(\k \cdot \hat{e}_3) - i n \pi} - 1}{(\k \cdot \hat{e}_3)^2 - (n\pi/L)^2},
\eeq
then the scattering amplitude reads
\beq
\mathcal{T}|_{\mathcal{O}(b^{2N})} = i E^*(\k; \iota) E(\k';\iota') \,  {\v}^{\dagger}(\k) \cdot {\bf T}^{N-1}(\omega) \cdot {\v}(\k'),
\eeq
and the resulting geometric series may be summed from $N=1$ to $\infty$ to get
\beq
\mathcal{T} =  {\v}^{\dagger}(\k) \cdot \frac{i  E^*(\k; \iota) E(\k';\iota')}{1-{\bf T}(\omega)} \cdot {\v}(\k')
\eeq
where $\dfrac{1}{1 - {\bf T}(\omega)}$ denotes the inverse operator of $(1 - {\bf T}(\omega))$. 

\section{Kinematical factors in the three-phonon scattering} \label{sec:kinematical}

Energy-momentum conservation dictates that in a scattering process of two massless particles, such as the phonons we have at hand, only a reduced number of processes are possible. Namely, given two incident phonons (with polarizations $\iota_1$ and $\iota_2$) and one outgoing phonon (with polarization $\iota$), it is necessary that $c_\iota > c_{\iota_1}$ or $c_\iota > c_{\iota_2}$ if non-collinear scattering is to happen.

Thus, we have three possible situations that will yield different kinematical restrictions between $\k$ and $\k_1$, and factors $K(\k,\k_1;\{c_{\iota_i}\}) \equiv \left| \frac{\partial (\omega_\iota - \omega_{\iota_1} - \omega_{\iota_2})}{\partial |\k|} \right|^{-1}$ as a consequence of performing the integral over possible outgoing momenta against the energy-conservation Dirac delta $\delta (\omega_\iota(\k) - \omega_{\iota_1}(\k_1) - \omega_{\iota_2}(\k + \k_1) )$. For all of these situations, let $\cos \theta \equiv \frac{\k_1 \cdot \k}{ |\k_1| |\k|} $.

\begin{enumerate}
\item $c_\iota = c_L$ and $c_{\iota_1} = c_{\iota_2} = c_T$:

Here we have
\beq
|\k| = \frac{2c_T (c_L - c_T \cos \theta)}{c_L^2 - c_T^2} |\k_1|,
\eeq
and
\beq
K(\k, \k_1; \{c_{\iota_i}\}) = \left| c_L - c_T^2 \frac{|\k| - |\k_1| \cos \theta }{c_L |\k| - c_T |\k_1|} \right|^{-1}.
\eeq

\item $c_{\iota_1} = c_\iota = c_L$ and $c_{\iota_2} = c_T$:

Here we have
\beq
|\k|^2 - 2|\k| |\k_1| \frac{c_L^2 - c_T^2 \cos \theta}{c_L^2 - c_T^2} + |\k_1|^2 = 0,
\eeq
and
\beq
K(\k, \k_1; \{c_{\iota_i}\}) = \left| c_L - c_T^2 \frac{|\k| - |\k_1| \cos \theta }{c_L (|\k| - |\k_1|)} \right|^{-1}.
\eeq

\item $c_{\iota_2} = c_\iota = c_L$ and $c_{\iota_1} = c_T$:

Here we have
\beq
|\k| = \frac{c_L^2 - c_T^2}{2c_L (c_L \cos \theta - c_T )} |\k_1|,
\eeq
and
\beq
K(\k, \k_1; \{c_{\iota_i}\}) = \left| c_L - c_L^2 \frac{|\k| - |\k_1| \cos \theta }{c_L |\k| - c_T |\k_1|} \right|^{-1}.
\eeq

\end{enumerate}

\section{Form Factors of the Quantum Corrections to the Scattering Amplitude } \label{sec:quant-form}

In section~\ref{sec:Quant}, we have defined and used three functions that characterize the form factors of the scattering amplitudes. The leading term includes the same factors as the quadratic theory:
\beq
f_n^{(0)}(k) \equiv n\pi  \frac{e^{ikL}(-1)^n - 1 }{(kL)^2 - (n\pi)^2},
\eeq
while the subleading terms (proportional to $\hbar$) also involve:
\begin{widetext}
\bea
f_{n,n'}^{(1)}(k) &\equiv& \frac{2i(n\pi)(n'\pi) (e^{ikL}(-1)^{n+n'} - 1)}{(kL)^4 + (n\pi)^4 + (n'\pi)^4 - 2(kL)^2 (n\pi)^2 - 2 (kL)^2 (n'\pi)^2 - 2(n\pi)^2 (n'\pi)^2 } \\
f_{n,n'}^{(2)}(k) &\equiv& \frac{n\pi  (e^{ikL}(-1)^{n} - 1)}{4} \left[ \frac{1}{(kL + 2\pi n')^2 - (n\pi)^2} + \frac{2}{(kL)^2 - (n\pi)^2} + \frac{1}{(kL - 2\pi n')^2 - (n\pi)^2} \right].
\eea
\end{widetext}

By inspecting these functions, it is fairly straightforward to verify that they have peaks at $kL \sim (n+n')\pi$ and at $kL \sim n\pi$ respectively. $f_{n,n'}^{(2)}(k)$ also has peaks at $kL \sim \pi(n \pm 2n')$, but they can be shown to be of smaller amplitude by evaluating the function at those points. Moreover, their peak values are of order $\mathcal{O}(1)$:
\bea
f_n^{(0)}(n\pi/L) &=& i/2 \\
f_{n,n'}^{(1)}((n+n')\pi/L) &=& -1/4 \\
f_{n,n'}^{(2)}(n\pi/L) &=& i/4.
\eea
Let us stress that these functions peak \textit{close} to those points in amplitude (which is why we have used the similar ``$\sim $'' symbol), and increasingly so as $n,n'$ take greater values.

Nonetheless, even though these are approximate features, the $\ln (n_D)$ scaling of the correction proportional to $\hbar$ will always be present as there is a term in $f_{n,n'}^{(2)}(k)$ that is exactly independent of $n'$, and has greater amplitude relative to the other terms.

\end{appendix}

\bibliographystyle{apsrev4-1_title.bst}
\bibliography{QD_bib.bib}

\end{document}